\documentclass[aps,prb,twocolumn,groupedaddress]{revtex4}

\usepackage[dvips]{graphicx}

\usepackage{amsmath}

\usepackage{amstext}

\usepackage{caption2} 
\usepackage{graphics}
\usepackage{exscale}
\usepackage{epsfig}
\usepackage{changebar}
\usepackage[latin1]{inputenc}
\usepackage{bm}
\usepackage{bbm}

\usepackage{rotating}
\usepackage{latexsym}

\usepackage{psfrag}

\renewcommand{\epsilon}{\varepsilon}

\newcommand{\partdera}[2]{\frac{\partial #2}{\partial #1}}

\newcommand{\Imag}{\mathrm{Im}}

\begin{document}

\title{Magnetic Field Effects on Quasiparticles in Strongly Correlated Local Systems}
\author{A.C. Hewson, J. Bauer and W. Koller}
\affiliation{Department of Mathematics, Imperial College, London SW7 2AZ. UK}
\date{\today} 
\begin{abstract}

We show  that quasiparticles in a magnetic field of arbitrary strength $H$ can be
described by field dependent parameters. We illustrate this approach in the 
case of an Anderson impurity model and use the numerical renormalization group (NRG) to
calculate the renormalized parameters for the levels with spin $\sigma$,
$\tilde\epsilon_{\mathrm{d},\sigma}(H)$, resonance 
width $\tilde\Delta(H)$ and the effective local quasiparticle interaction
$\tilde U(H)$. In the Kondo or strong correlation
limit of the model the progressive de-renormalization of the quasiparticles can be
followed as the magnetic field is increased. The low temperature behaviour,
including the conductivity, in
arbitrary
magnetic field can be calculated  in terms of the field dependent parameters
using the renormalized perturbation expansion. Using the NRG the field dependence of the
spectral density on higher scales  is also calculated.   
 \end{abstract}
\maketitle

\section{ Introduction}\par
Electrons in  strongly correlated systems are particularly sensitive to the
application of  magnetic fields. One reason is that strong correlations are
usually a consequence of the interaction of electrons with enhanced spin
fluctuations, and these fluctuations couple strongly to a magnetic
field. Another reason is that there is a low temperature scale $T^*$ ($T^*\ll
T_{\rm F}$) induced which plays the role of an effective 
Fermi temperature $T_{\rm F}$. The effects of a magnetic
field $H$ in general depend on the ratio of the two energy scales  $\mu_{\rm B}H$ and
$k_{\rm B}T_{\rm F}$. In a weakly correlated metal 
 $\mu_{\rm B}H/k_{\rm B}T_{\rm F}\ll 1$, but in a strongly
correlated system the relevant ratio is $\mu_{\rm B}H/k_{\rm B}T^*$, which
can be of order unity.  This sensitivity 
means  that a magnetic field is an important  tool in the experimental
investigation of strongly correlated  metallic systems, such as magnetic
impurities, quantum dots, heavy fermions and transition metal oxides.

In this paper we concentrate mainly on the effects of a magnetic field on the
quasiparticles in the Fermi liquid regime for various models 
of strongly correlated metals. In particular we develop an approach, based on
a combination of the numerical 
renormalization group (NRG)\cite{Wil75,KWW80a} and renormalized perturbation theory
(RPT)\cite{Hew93,Hew01}. This approach  gives a comprehensive picture of
the behaviour of quasiparticles in magnetic fields of arbitrary strength, such
that we can follow the renormalization 
or de-renormalization of the quasiparticles as the magnetic field strength is
changed. In this paper we concentrate on developing the approach 
 for the single impurity Anderson model (SIAM), as a model of
magnetic impurities and quantum dots. This prepares the way for
generalizing the approach to lattice models for transition metal oxides and
heavy fermion materials, using dynamical mean field theory
(DMFT)\cite{GKKR96}, which will be the subject of a subsequent paper. \par 

\section{ Quasiparticles at $T=0$ with Zero Magnetic Field}
The generic model for strongly correlated local systems, such as magnetic
impurities in a host metal or a quantum dot coupled to an electron reservoir,
is the Anderson model \cite{And61}. The Hamiltonian for this model is
\begin{eqnarray}
&&H_{\rm AM}=\sum\sb {\sigma}\epsilon\sb {\mathrm{d},\sigma} d\sp {\dagger}\sb
{\sigma}  
d\sp {}\sb {\sigma}+Un\sb {\mathrm{d},\uparrow}n\sb {\mathrm{d},\downarrow} \label{ham}\\
&& +\sum\sb {{ k},\sigma}( V\sb { k}d\sp {\dagger}\sb {\sigma}
c\sp {}\sb {{ k},\sigma}+ V\sb { k}\sp *c\sp {\dagger}\sb {{
k},\sigma}d\sp {}\sb {\sigma})+\sum\sb {{
k},\sigma}\epsilon\sb {{ k},\sigma}c\sp {\dagger}\sb {{ k},\sigma}
c\sp {}\sb {{
k},\sigma}, \nonumber
\end{eqnarray}
where $\epsilon_{\mathrm{d},\sigma}=\epsilon_{\rm d}-\sigma g\mu_{\rm B} H/2$
is the energy of the localized  level at an impurity site or 
quantum dot in a magnetic field $H$, $U$ the interaction at this local site,
 and $V_{k}$ the hybridization matrix element to a band of conduction electrons with
energy $\epsilon_k$. When $U=0$ the local level broadens into a resonance, 
corresponding to a localized quasi-bound state, whose width depends on
the quantity $ \Delta(\omega)=\pi\sum\sb {k}| V\sb {k}|\sp 2\delta(\omega
-\epsilon\sb { k})$. It is usual to consider the case 
of a wide conduction band with a flat density of states where
$\Delta(\omega)$ becomes independent of $\omega$ and can be taken as a
constant $\Delta$.\par 
The low energy behaviour of this model can be expressed in terms of the
renormalized quasiparticles of a local Fermi liquid, which is described by 
a renormalized version of the same model:
\begin{eqnarray} 
&&\tilde H_{\rm AM}=\sum\sb {\sigma}\tilde\epsilon\sb {\mathrm{d}}
d\sp {\dagger}\sb {\sigma}
d\sp {}\sb {\sigma}+
\tilde U : n\sb {\mathrm{d},\uparrow}n\sb {\mathrm{d},\downarrow}: \nonumber \\
&&+\sum\sb {{ k},\sigma}(\tilde V\sb { k}d\sp {\dagger}\sb {\sigma}
c\sp {}\sb {{ k},\sigma}+\tilde V\sb { k}\sp *c\sp {\dagger}\sb {{
k},\sigma}d\sp {}\sb {\sigma})+\sum\sb {{
k},\sigma}\epsilon\sb {{ k},\sigma}c\sp {\dagger}\sb {{ k},\sigma}
c\sp {}\sb {{
k},\sigma},\label{rham}
\end{eqnarray}
where  the colon brackets indicate that the expression within them must be
normal-ordered.  
This Hamiltonian corresponds to the low energy fixed point of the Wilson
numerical renormalization group 
transformation of the discretized Anderson and Kondo models, with the leading
irrelevant terms \cite{Wil75,KWW80a,Hew93b}. The advantage of describing the fixed
point in this way, as a renormalized  Anderson model rather than as a strong coupling
fixed point of the Kondo model, even in the strong correlation or Kondo limit,
is that it clearly brings out the 1-1 correspondence of the low-lying single
particle excitations with those of the non-interacting model
\cite{HOM04,Hew93b,Hew05}. Furthermore, it is applicable in all parameter regimes, from
weak to strong coupling and for all occupation values for the local site.
The parameters $\tilde\epsilon_{\mathrm{d}}$, $\tilde\Delta$, and $\tilde U$
define the quasiparticles of 
this renormalized model, and a simple direct procedure for calculating these
parameters  using the NRG has been given  earlier \cite{HOM04,Hew05}.
In terms of these parameters the Friedel sum rule \cite{Fri56,Lan66}, which
gives the total occupation of the  d-orbital at the impurity
site $n_{\mathrm{d}}$ for $T=0$ in the wide band limit, is 
\begin{equation}  
n_{\mathrm{d}}={1}-{2\over\pi}\tan ^{-1}\left
({\tilde\epsilon_{{\rm d}}}\over{\tilde\Delta}\right
).\label{qpfsr}
\end{equation} 
In the wide band limit the renormalized parameters, $\tilde\epsilon_{\mathrm{d}}$,
$\tilde\Delta$, and $\tilde U$, can be expressed as functions of two
independent 'bare' parameters, which can be taken to be $U/\pi\Delta$ and
$\epsilon_{\mathrm{d}}/\pi\Delta$. Some typical plots of the renormalized
parameters as a functions of these variables were presented in the earlier work \cite{HOM04}.
An alternative way of presenting the results is in terms of just one of these
variables,  $U/\pi\Delta$, 
and the occupation of the impurity levels $n_{\mathrm{d}}$, which gives a global picture
over the various regimes of the model. \\

\begin{figure}
\begin{center}
\vskip.4cm
\includegraphics[width=0.4\textwidth,height=5cm]{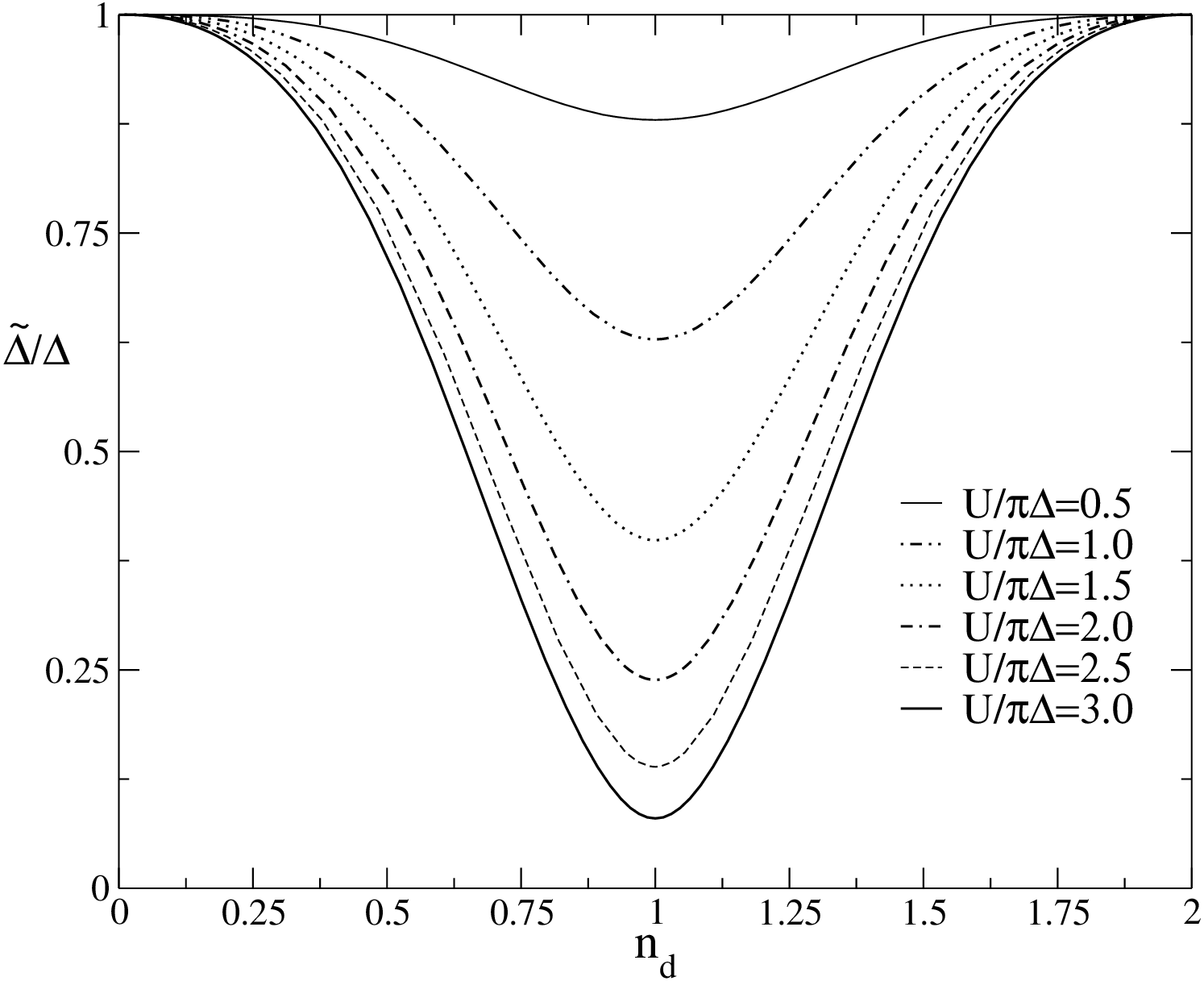}\\
\centerline {(a)}
\vskip.2cm
\includegraphics[width=0.4\textwidth,height=5cm]{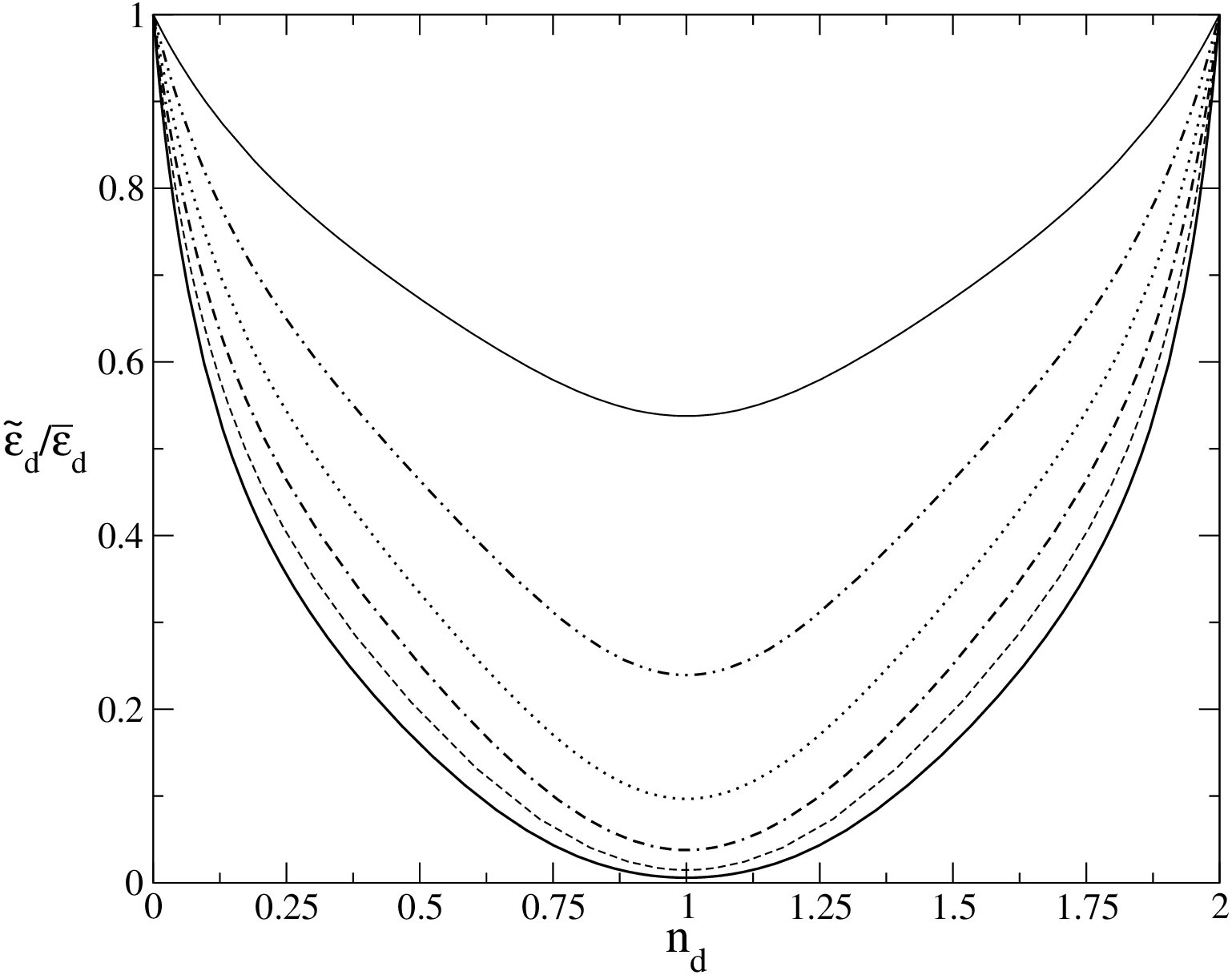}\\
\centerline {(b)}
\vskip.2cm
\includegraphics[width=0.4\textwidth,height=5cm]{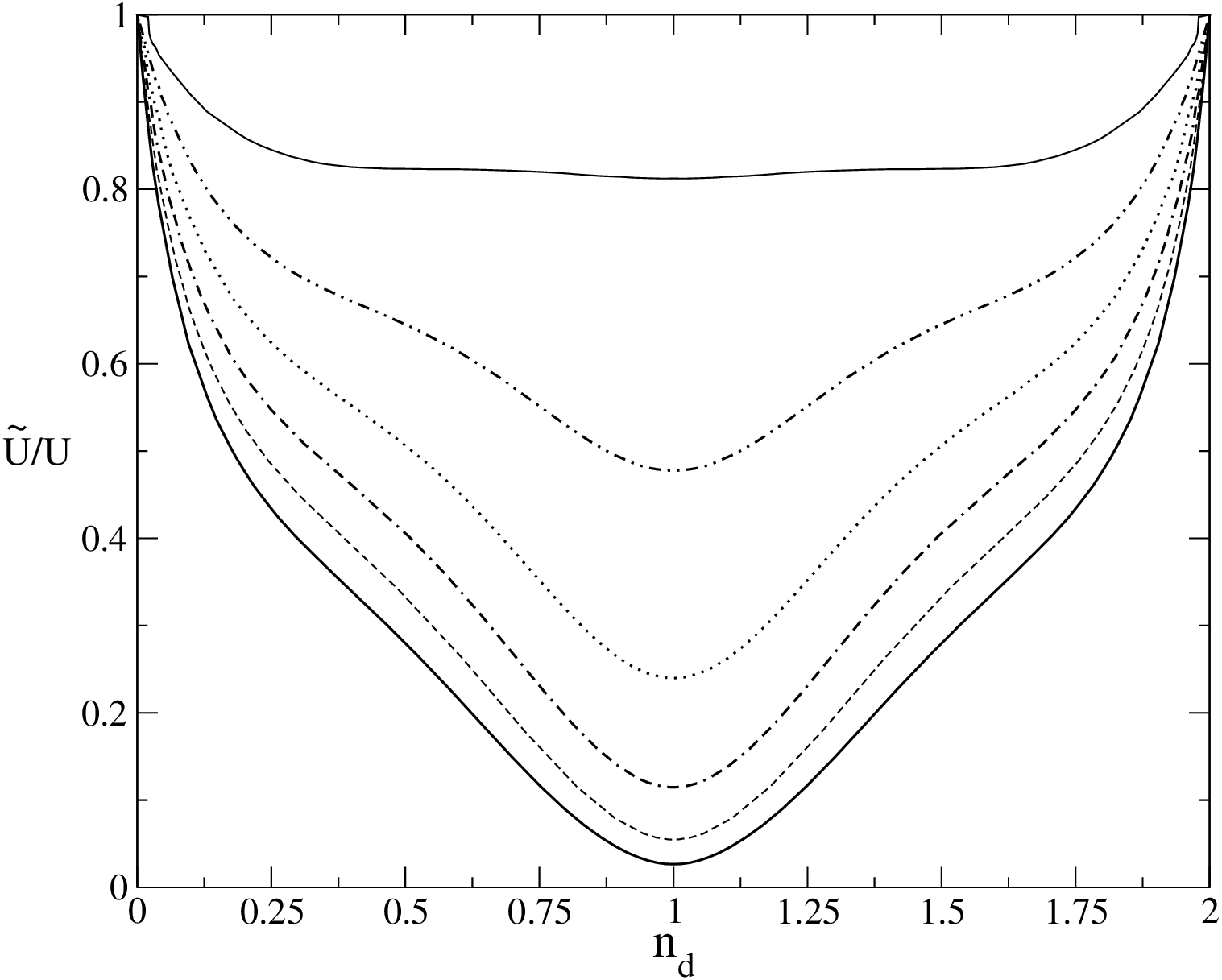}\\
\centerline {(c)}
\vspace{-0.2cm}
\label{figure1}
\end{center} 
\caption{Plots of the renormalized parameters for the Anderson model, (a)
  $\tilde \Delta/\Delta$, 
(b) $\tilde\epsilon_{\mathrm{d}}/\bar\epsilon_{\mathrm{d}}$  and (c) $\tilde U/U$, as a function
of the occupation of the impurity site $n_{\mathrm{d}}$, for values of 
$U/\pi\Delta=0.5,1.0,1.5,2.0,2.5,3.0$.}
\end{figure}

In figures 1(a), 1(b) and 1(c) we present results for the ratios,
$\tilde\Delta/\Delta$, $\tilde\epsilon_{\mathrm{d}}/\bar\epsilon_{\mathrm{d}}$ 
and $\tilde U/U$, where $\bar\epsilon_{\mathrm{d}}=\epsilon_{\mathrm{d}}+U/2$,
which give a measure of the degree of renormalization of 
these parameters for the 'bare' values of $U/\pi\Delta=0.5,1,1.5,2,2.5,3$. 
In the empty ($n_{\mathrm{d}}\to 0$) and full ($n_{\mathrm{d}}\to 2$) regimes
all the parameters approach their unrenormalized values 
so these ratios tend to unity. The renormalizations are the most pronounced
in the region $n_{\mathrm{d}}\simeq 1$. In the almost localized (Kondo)
regime, $n_{\mathrm{d}}\simeq 1$, $U/\pi\Delta>2$, there is only one energy
scale, the Kondo temperature 
$T_{\rm K}$, such that $\tilde\epsilon_{\mathrm{d}}\simeq 0$ 
and $\tilde U=\pi\tilde\Delta=4T_{\rm K}$.\par The impurity density of states
for the non-interacting quasiparticles $\tilde\rho_{\mathrm{d}}(\omega)$ at $T=0$ is
given by 
\begin{equation}
\tilde\rho_{\mathrm{d}}(\omega)={\tilde\Delta/\pi\over
  (\omega-\tilde\epsilon_{\mathrm{d}})^2+\tilde\Delta^2},
\label{qpdos}
\end{equation} 
and in the localized regime corresponds to a Kondo resonance of half-width
$\tilde\Delta=4T_{\rm K}/\pi$
at the Fermi level, where the Kondo temperature is defined in terms of the
impurity spin susceptibility $\chi_s$ at $T=0$ via $T_{\rm K}=(g\mu_{\rm  B})^2/4\chi_s$.\par 
As in the general Fermi liquid theory, the linear coefficient of the specific heat
for the impurity can be expressed simply in terms of the quasiparticle density
of states at the Fermi level, and is given by  
\begin{equation}
\gamma_{\mathrm{d}}={2\pi^2\over 3}\tilde\rho_{\mathrm{d}}(0).\label{rgam}
\end{equation}
Exact expressions for impurity spin and charge susceptibilities, $\chi_s$ and
$\chi_c$, in terms of the renormalized parameters for $T=0$ are  
\begin{equation}
\chi_{s}={1\over 2}\tilde\rho_{\mathrm{d}}(0)(1+
 \tilde U\tilde\rho_{\mathrm{d}}(0)),\quad\chi_{c}={1\over 2}\tilde
 \rho_{\mathrm{d}}(0)(1-\tilde
 U\tilde\rho_{\mathrm{d}}(0)).\label{rsus}
\end{equation}
These results can be derived using the renormalized perturbation approach\cite{Hew93,Hew01}
working simply only up to first order in $\tilde U$  (see also the Appendix
here).\par
It is interesting to note that these Fermi liquid results apply even in the limit
$\epsilon_d\to -\infty$, $U\to \infty$, when the charge fluctuations
are completely suppressed and the Kondo scale $T_{\rm K}\to 0$.
In this limit $\tilde U\tilde\rho_d(0)\to 1$ and
$\tilde\rho_d(\omega)\to\delta(\omega)$ corresponding to the low energy excitations
of an isolated spin. If the finite temperature Fermi distribution function is
included in the calculations leading to  $\chi_c$ and $\chi_s$ in equation 
(\ref{rsus}), then in this limit $\chi_c\to 0$, and $\chi_s\to (g\mu_{\rm
B})^2/4k_{\rm B}T$, the Curie law for a localized magnetic moment.\par

So far we have assumed a repulsive interaction $U>0$, but the formulae are
equally applicable to the case of an attractive interaction $U<0$. In this
case, however, except at the particle-hole symmetric point, the renormalized
parameters behave quite differently as a function of $|U|/\pi\Delta$ and  $n_{\mathrm{d}}$. In
 figures 2(a), 2(b) and 2(c), we present results for the ratios,
 $\tilde\Delta/\Delta$, $\tilde\epsilon_{\mathrm{d}}/\bar\epsilon_{\mathrm{d}}$, 
and $\tilde U/U$,  for values of $U/\pi\Delta=-(0.5,1,1.5,2,2.5,3)$. 

\begin{figure}
\begin{center}
\vskip0.4cm
\includegraphics[width=0.4\textwidth,height=5cm]{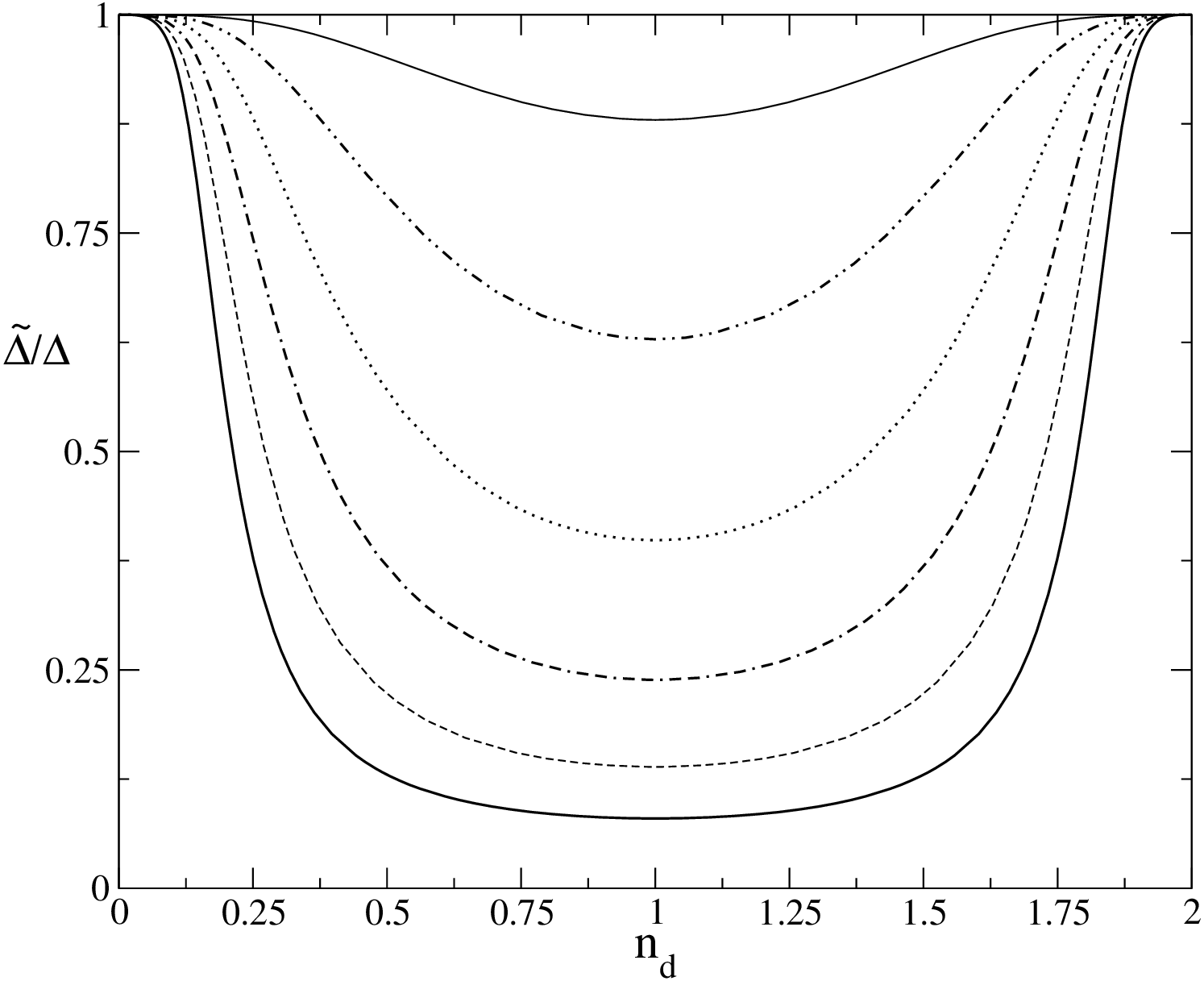}\\
\centerline {(a)}
\vskip.2cm
\includegraphics[width=0.4\textwidth,height=5cm]{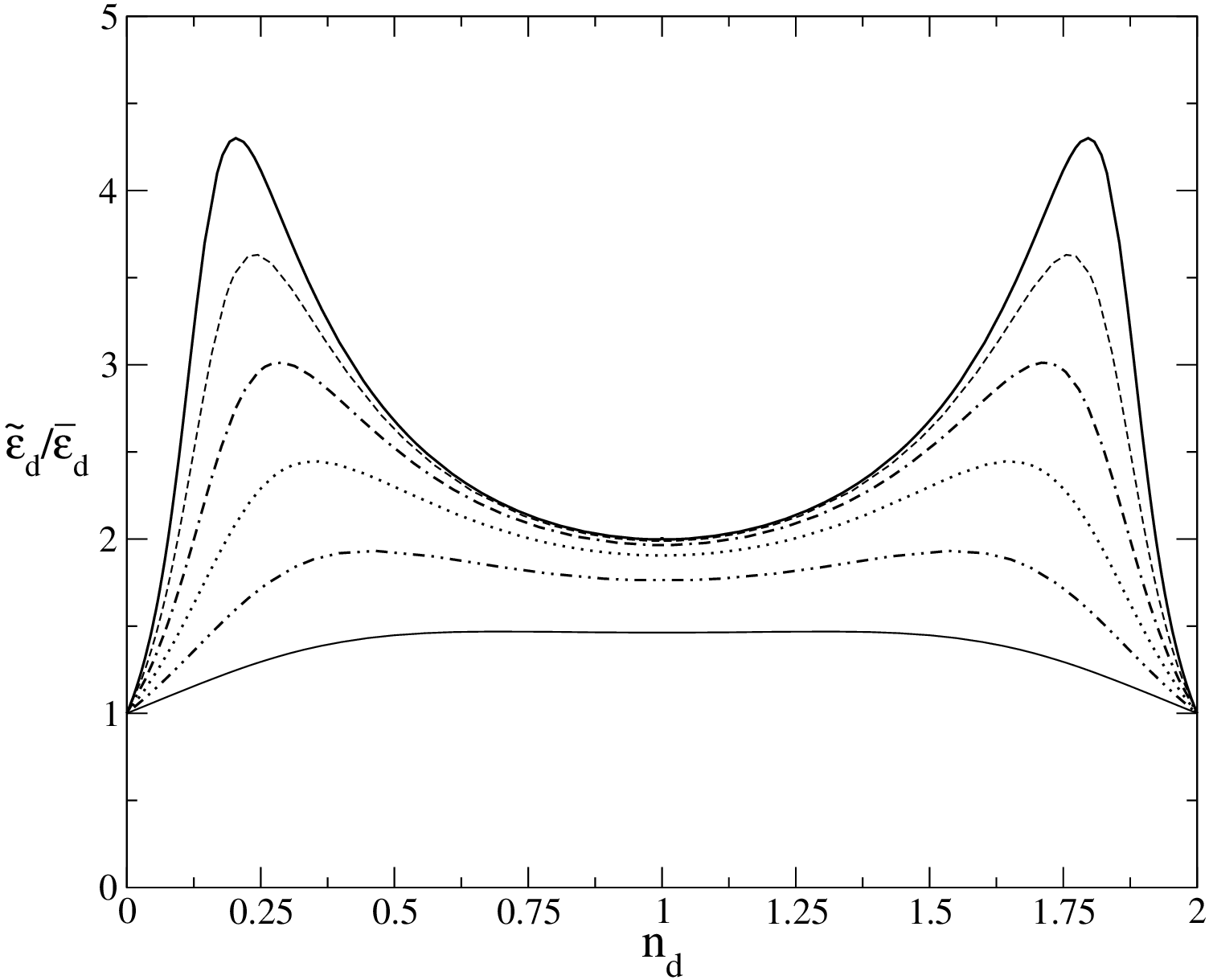}\\
\centerline {(b)}
\vskip.2cm
\includegraphics[width=0.4\textwidth,height=5cm]{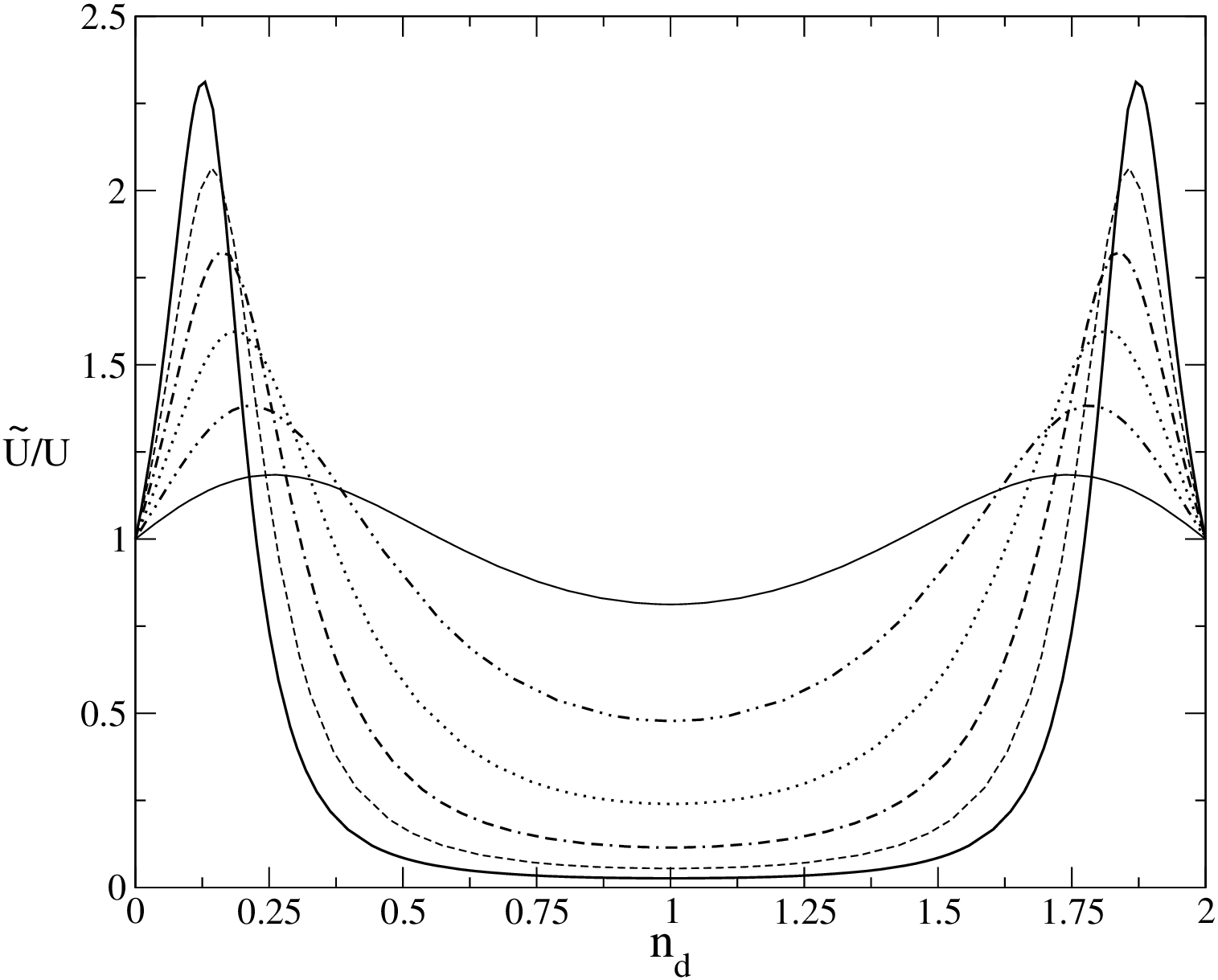}\\
\centerline {(c)}
\vskip.2cm
\label{figure2}
\end{center} 
\vspace{-0.3cm}
\caption{Plots of the renormalized parameters for the Anderson model, (a)
  $\tilde \Delta/\Delta$, (b)
  $\tilde\epsilon_{\mathrm{d}}/\bar\epsilon_{\mathrm{d}}$  and (c) $\tilde
  U/U$, as a function of the occupation of the impurity site $n_{\mathrm{d}}$,
  for values of  $U/\pi\Delta=-(0.5,1.0,1.5,2.0,2.5,3.0)$.}
\end{figure} 
In comparing the results in figure 2(a)(b) and (c) with the corresponding
results in figure 1 (a)(b) and (c), it can be seen that, irrespective of the
sign of $U$ all the parameters converge to their bare values in limits $n_{\mathrm{d}}\to
0$ and $n_{\mathrm{d}}\to 2$ and the values of $\tilde\Delta/\Delta$ and 
$\tilde U/U$  are the same at the symmetric point $n_{\mathrm{d}}=1$, they are remarkably
different elsewhere. In the range $0.5<n_{\mathrm{d}}<1.5$,  $\tilde\Delta/\Delta$ and
$\tilde U/U$  for $U<0$ are remarkably flatter than the corresponding results
for $U>0$, especially for large values of $|U|$.  The ratios
$\tilde\epsilon_{\mathrm{d}}/\bar\epsilon_{\mathrm{d}}$ 
and  $\tilde U/U$  for $U<0$ do not increase monotonically with increase of
$|n_{\mathrm{d}}-1|$, 
as they do in the corresponding  $U>0$ results.
The ratio $\tilde U/U$  develops a sharp peak with values such that $\tilde 
U>U$ and is completely different from the positive $U$ counterpart.
The ratio $\tilde\epsilon_{\mathrm{d}}/\bar\epsilon_{\mathrm{d}}$ is also
completely different and is greater than, or equal to, unity over the whole
range, developing a similar peak to that for  $\tilde U/U$.  At the particle-hole
symmetric point,  $n_{\mathrm{d}}=1$,   $\bar\epsilon_{\mathrm{d}}=0$, and
$\tilde\epsilon_{\mathrm{d}}=0$, so the ratio
$\tilde\epsilon_{\mathrm{d}}/\bar\epsilon_{\mathrm{d}}$  is not defined. In
the positive $U$ case, the ratio tends to 0 for large $|U|$ and to 2 in the
negative $U$ case.\par 
The main applications of the Anderson model to magnetic impurities and quantum
dots are for the positive $U$ case, though the negative $U$ model 
has some limited application as an effective model for some locally coupled
electron-phonon systems,  where a local attraction is induced through the
exchange of a virtual phonon  \cite{HM02}. However, in the next section
we shall exploit the negative $U$ model here by using  spin-isospin symmetry
to transform it into a model with positive $U$ with particle-hole symmetry in
the presence of a magnetic field. The restriction 
to the  $n_{\mathrm{d}}=1$ particle-hole symmetric case will not be a serious
limitation because in the strong correlation/local moment limit the
renormalized level rapidly approaches the  Fermi level for large $U$,
$\tilde\epsilon_{\mathrm{d}}\to 0$, and hence via the Friedel sum rule
$n_{\mathrm{d}}\to 1$. In this particle-hole symmetric case there is only one  
energy scale for a given $U/\pi\Delta$, which we denote by $T^*$ and define by
 $T^*=\pi\tilde\Delta/4$; it is such that $T^*\to T_{\rm K}$ in the Kondo
 regime $U/\pi\Delta>2$. \par

\section{Quasiparticles at $T=0$ with Arbitrary Magnetic Field $H$}
The spin-isospin transformation which formally eliminates the magnetic field
term in the symmetric model 
 is defined by
\begin{equation} c^{\dagger}_{\mathrm{d}\uparrow}\to  c^{\dagger}_{\mathrm{d}\uparrow},\quad
  c^{\dagger}_{\mathrm{d}\downarrow}\to c^{}_{\mathrm{d}\downarrow}
\end{equation}

and 
 \begin{equation} c^{\dagger}_{k\uparrow}\to  c^{\dagger}_{k\uparrow},\quad
  c^{\dagger}_{k\downarrow}\to c^{}_{-k\downarrow}
\end{equation}
with the requirement $\epsilon_{-k}=-\epsilon_{k}$ with $V^*_{-k}=- V_{k}$.
With this transformation for the symmetric model with $U>0$ and
$\bar\epsilon_{\mathrm{d}}=0$ in the presence of a magnetic field $H$ 
at the impurity site gets mapped into the model with an interaction $-U$, and
no magnetic field but with $\bar\epsilon_{\mathrm{d}}=g\mu_BH/2=h$. 
The spin/charge susceptibilities for the positive $U$ symmetric model in a
magnetic field are then given by the charge/spin susceptibilities  of the
model with a negative $U$. The magnetic field dependent parameters
$\tilde\Delta(h)$, $\tilde\epsilon_{\mathrm{d},\sigma}(h)$  and $\tilde U(h)$,
for the $U>0$ model  then correspond to  $\tilde\Delta(h)$,
$-\sigma\tilde\epsilon_{\mathrm{d}}(h)$  and $-\tilde U(h)$ for the
corresponding negative $U$ model with $\bar\epsilon_{\mathrm{d}}=h$. The
Friedel sum rule is still applicable to each spin component  from which the
induced impurity magnetization $M(h)$  at $T=0$ can be deduced,
\begin{equation} 
m(h)={M(h)\over g\mu_{\rm B}}={1\over
    2}(n_{\mathrm{d},\uparrow}-n_{\mathrm{d},\downarrow})={1\over\pi}{\rm
    tan}^{-1}\left({\tilde\epsilon_{\mathrm{d}}(h)\over\tilde\Delta(h)}\right),
\label{mag}
\end{equation}
from the two parameters $\tilde\epsilon_{\mathrm{d}}(h)$ and $\tilde\Delta(h)$
that characterize the non-interacting quasiparticles. 
The quasiparticle density of states for the spin up and spin down electrons is
given by
\begin{equation}
\tilde\rho_{\mathrm{d},\sigma}(\omega,h)={\tilde\Delta(h)/\pi\over
  (\omega-\sigma\tilde\epsilon_{\mathrm{d}}(h))^2+\tilde\Delta^2(h)},
\label{hqpdos}
\end{equation}
and the field dependent spin and charge susceptibilities at $T=0$ are given by
$$\chi_{s}(h)={1\over 2}\tilde\rho_{\mathrm{d}}(0,h)(1+
 \tilde U(h)\tilde\rho_{\mathrm{d}}(0,h)),$$
\begin{equation}
\chi_{c}(h)={1\over 2}\tilde
 \rho_{\mathrm{d}}(0,h)(1-\tilde
 U(h)\tilde\rho_{\mathrm{d}}(0,h)).\label{hrsus}\end{equation}
As $\tilde\rho_{\mathrm{d},\sigma}(0,h)$ is independent of the spin state
we can drop the spin index $\sigma$.\par
   As $\tilde\epsilon_{\mathrm{d},\sigma}(h)$ is entirely  magnetic field driven it is convenient to write it as $\tilde\epsilon_{\mathrm{d},\sigma}(h)=h\tilde\eta(h)$, then $2h\tilde\eta(h)$ is the Zeeman splitting of the impurity levels for the non-interacting quasiparticles.\par 

Using these results we can reinterpret the renormalized parameters in figures
2(a), (b) and (c) for the $U<0$ model as $\tilde\Delta(h)/\Delta$, 
$\tilde\eta(h)$ and $\tilde U(h)/U$ for the $U>0$ model as functions of the
variable $2m(h)+1$, which replaces the $n_{\mathrm{d}}$. The fact that
$\tilde\epsilon_{\mathrm{d}}/\bar\epsilon_{\mathrm{d}}$ in figure 2(b)
approaches the value 2 as $n_{\mathrm{d}}\to 1$ in the strong correlation
limit can now be interpreted as $\tilde\eta(h)\to 2$ for $h\to 0$, 
which is equivalent to the  Wilson ratio $R=4\pi\chi_s/3(g\mu_{\rm
  B})^2\gamma_{\rm imp}=2$ for the strongly renormalized quasiparticles 
in zero field, which is enhanced  compared with  the free electron value
$R=1$. In figure 2(c) the sharp rise  in $\tilde U/U$ on reducing
$n_{\mathrm{d}}$ from the value $n_{\mathrm{d}}=2$, can be interpreted as the
enhancement of the effective interaction $\tilde U(h)$, as the magnetization
is reduced from the saturated value, $m_{\rm sat}=1/2$. As the applied
magnetic field is reduced from the regime  $h>U$, spin fluctuations increase
and enhance the effective interaction $\tilde U$, as in the random phase
approximation (RPA), above the bare value $U$ (we will see this more
explicitly later).  As the magnetic field is further reduced the many-body
correlations are increasingly effective in screening the impurity so that
$\tilde U(h)$ decreases from an enhanced value greater than $U$ to a value
$4T_{\rm K}$ as $h\to 0$ when $U>2\pi\Delta$. 

\vspace{0.8cm}
\begin{figure}[!htb]
\begin{center}
\includegraphics[width=0.42\textwidth]{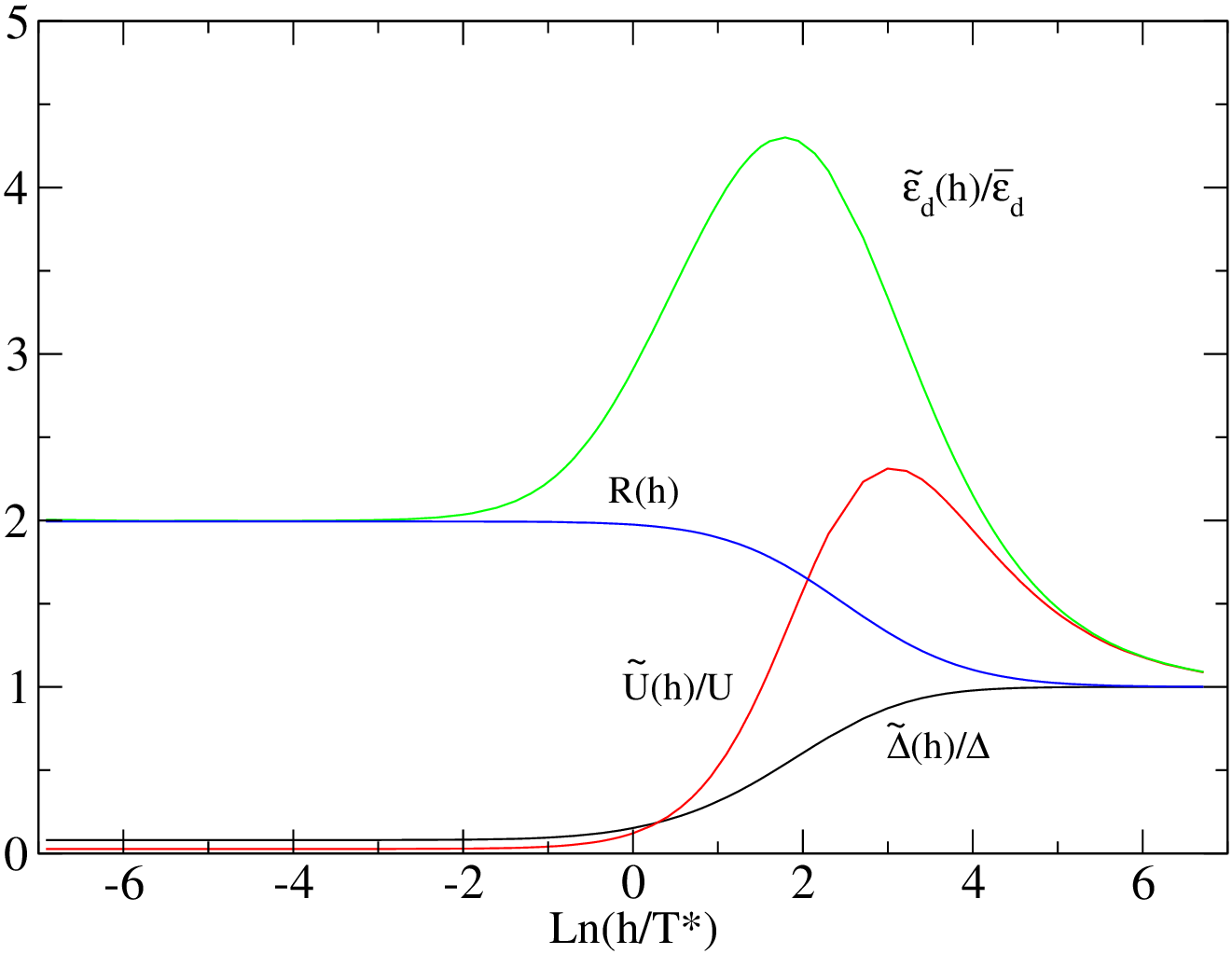}
\label{figure3}
\end{center}
\caption{The magnetic field dependence of the 
renormalized parameters  $\tilde \Delta(h)/\Delta$, 
$\tilde\epsilon_{\mathrm{d}}(h)/\bar\epsilon_{\mathrm{d}}$ ($=\tilde\eta(h)$)  
 and $\tilde U(h)/U$, and the Wilson ratio $R(h)$, for the symmetric  Anderson
 model with $U /\pi\Delta=3.0$ plotted on a logarithmic scale.}
\vskip.6cm
\end{figure}

In figure 3 we give a more conventional plot of the renormalized parameters 
 as a function of the natural logarithm of the magnetic field, ${\rm
 ln}(h/T^*)$, for the strong coupling case $U/\pi\Delta=3$,  where
 $T^*=\pi\tilde\Delta(0)/4=2.00\times 10^{-2}=T_{\rm K}$, in agreement to
 within 0.5\% with  formula\cite{HZ85} for $T_{\rm K}$ for this model, $T_{\rm
 K}=\sqrt{(U\Delta/2)}e^{-\pi U/8\Delta+\pi\Delta/2U}$.  
We can follow the progressive de-renormalization of the quasiparticles as the
 strong correlation effects are suppressed as the magnetic field is
 increased. Initially 
 the quasiparticle interaction $\tilde U(h)$ increases and can reach values
 such that $\tilde U(h)$ is greater than the bare interaction $U$.
This does not imply, however, that the interaction effects are becoming
 stronger. The effects of the interaction on the low energy scale 
depend upon the combination, $\tilde U(h)\tilde\rho_{\mathrm{d}}(0,h)$, and
$\tilde\rho_{\mathrm{d}}(0,h)$ falls off rapidly with $h$ as
 $\tilde\epsilon_{\mathrm{d}}(h)$ moves away from the Fermi level.   
The Wilson ratio $R(h)=1+\tilde U(h)\tilde\rho_{\mathrm{d}}(0,h)$ is a measure of this 
combination of factors and for the Kondo model it is known from 
Bethe ansatz calculations \cite{TW83} that $R(h)=2$ is independent of $h$. 
This can be seen to be the case in the results for $R(h)$ shown in figure 3
when the parameters correspond to the localized or Kondo regime.
The localized model, however, is only valid when the charge fluctuations are
completely suppressed.
For very large field values $h>U$ 
local charge fluctuations can be induced by the magnetic field and, as this
regime is approached, $R(h)$  makes a crossover to the value $R=1$
for non-interacting electrons.
The combination $\tilde
U(h)\tilde\rho_{\mathrm{d}}(0,h)$
can be seen to decrease monotonically with increase of $h$.
\par
\vskip.8cm
\begin{figure}
\begin{center}
\includegraphics[width=0.42\textwidth]{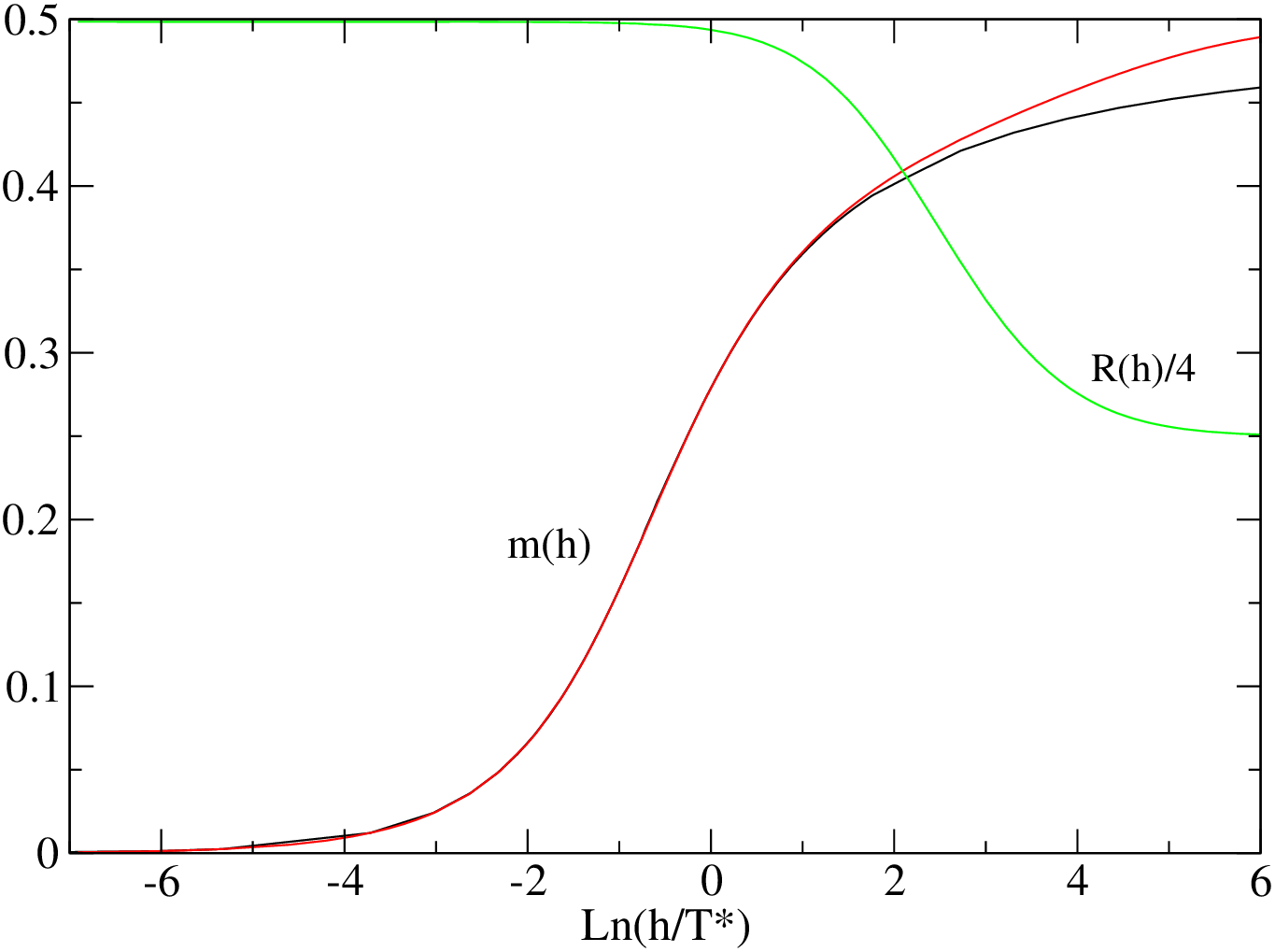}
\smallskip
\end{center}
\caption{The impurity magnetization $m(h)$  for the symmetric model with
  $U /\pi\Delta=3.0$, together with $R(h)/4$, where $R(h)$ is the
 Wilson ratio, plotted as a function of the logarithm of the
 magnetic field. Also shown for comparison are the corresponding Bethe ansatz results \cite{TW83} for the
field induced magnetization for the  
Kondo model. }
\label{figure4}
\vskip.4cm
\end{figure}
Equation  (\ref{hrsus}) for the susceptibility
$\chi_s(h)$ has a term in $\tilde U(h)$. However, the susceptibility can also be
derived by differentiating the expression (\ref{mag}) for the magnetization
which depends explicitly  only on the variables   $\tilde
\epsilon_{\mathrm{d}}(h)$ and  $\tilde \Delta(h)$. Hence the value of $\tilde
U(h)$ is not independent of the other two parameters and we can derive  a
relation between them, 
\begin{equation} 1+\tilde
  U(h)\tilde\rho_{\mathrm{d}}(0,h)={\partial\tilde\epsilon_{\mathrm{d}}(h)\over\partial
  h}-{\tilde\epsilon_{\mathrm{d}}(h)\over\tilde\Delta(h)}{\partial\tilde\Delta(h)\over\partial
  h}.\label{wi1}\end{equation}
The proof that  equation  (\ref{hrsus}) for the susceptibility  is exact
  depends on a Ward identity, so the relation (\ref{wi1}) we have derived must
  be an alternative statement of this identity. In terms of
  $\tilde\eta(h)=\tilde\epsilon_{\mathrm{d}}(h)/h$ it becomes
\begin{equation} 1+\tilde
  U(h)\tilde\rho_{\mathrm{d}}(0,h)=\tilde\eta(h)+h{\partial\tilde\eta(h)\over\partial
  h}-{h\tilde\eta(h)\over\tilde\Delta(h)}{\partial\tilde\Delta(h)\over\partial
  h}.\label{wi2}\end{equation}

For $h=0$ it implies that $\tilde\eta(0)=R(0)$, which can be seen in the
  results in figure 3. \par
Mean field theory can also be interpreted in terms of renormalized
  parameters with
$\tilde\epsilon_{\mathrm{d}}(h)=h+Um_{\rm MF}(h)$, $\tilde\Delta(h)=\Delta$,
where  $m_{\rm MF}(h)$ is the mean field magnetization.
 These parameters are 
  substituted
into equation (\ref{mag}),  and  $m_{\rm MF}(h)$
is determined self-consistently. From equation (\ref{wi1}) we can deduce the
  corresponding
value of $\tilde U(h)$ which gives
\begin{equation} \tilde U(h)={U\over {1-U\tilde\rho_{\mathrm{d}\rm
  MF}(0,h)}},\end{equation}
where $ \tilde\rho_{\mathrm{d}\rm MF}(\omega,h)$ is the quasiparticle density of states with
the mean field parameters.
This result corresponds to the enhancement of the susceptibility that  one
 finds from the random phase approximation. If the magnetic field is reduced
from a large value $h>U$ then $\tilde\rho_{\mathrm{d}\rm MF}(0,h)$ decreases and so $\tilde
 U(h)$ increases. This is precisely what is seen in the large $h$ regime in
the results in figure 3, as well as those shown earlier in figure 2 (b).\par

In figure 4 we plot the magnetization derived from these parameters for $U/\pi\Delta=3$ as
a function of ${\rm ln}(h/T^*)$ using equation (\ref{mag}). 
We also give the magnetization deduced from the Bethe ansatz results for the Kondo model for comparison together
with the value of $R(h)/4$. There is complete agreement with the results of
the Kondo model, up to and just beyond the point  
at which local charge fluctuations are induced by the magnetic field, where  
$R(h)/4$ begins to decrease significantly from its strong correlation value
0.5. The ratio $\tilde\epsilon_{\mathrm{d}}(h)/\tilde\Delta(h)$ must be a universal
function of $h/T_{\rm K}$ in the Kondo regime. For $h>T_{\rm K}$  we have
the asymptotic form $1/2-m(h)\sim 1/4{\rm ln}(h/T_{\rm K})$ \cite{AFL83}, so from 
equation (\ref{mag}), in this regime the ratio
$\tilde\epsilon_{\mathrm{d}}(h)/\tilde\Delta(h)\propto {\rm ln}(h/T_{\rm K})$. In figure
5 we plot the ratio  $\tilde\epsilon_{\mathrm{d}}(h)/\tilde\Delta(h)$ against ${\rm
ln}(h/T^*)$, for $U/\pi\Delta=5$,  
which is well in the Kondo regime so $T^*=T_{\rm K}$. It can be seen this
ratio is proportional to ${\rm ln}(h/T_{\rm K})$  for a significant range of
magnetic field values for $h>T_{\rm K}$, before the effects of the field
induced charge fluctuations take over. Due to the charge fluctuations, the
approach to saturation is much more rapid for the Anderson model than for the
Kondo model, once $h$ exceeds $U$, as can be seen clearly in figure 4.\par 

\begin{figure}[htb!]
\begin{center}
\vskip0.4cm
\includegraphics[width=0.4\textwidth,height=5cm]{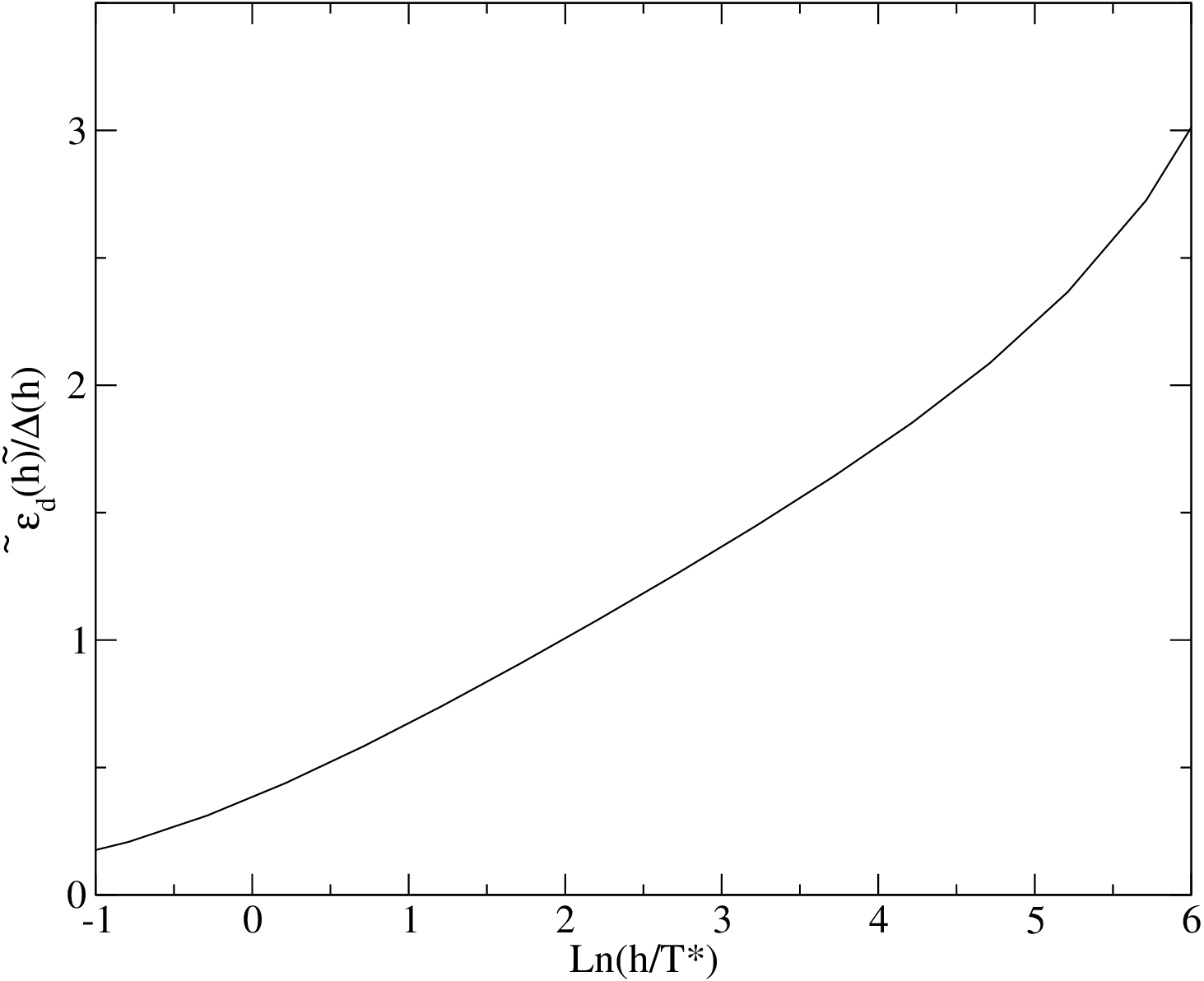}
\smallskip
\label{figure5}
\end{center}
\caption{The ratio $\tilde\epsilon_{\mathrm{d}}(h)/\tilde\Delta(h)$
for  $U/\pi\Delta =5.0$ plotted as a function of the logarithm of the
 magnetic field.}
\vskip.4cm
\end{figure}

A  similar plot of the magnetization $m(h)$ against field on a logarithmic
scale in shown in figure 6 for a weak correlation case
$U/\pi\Delta=0.25$. In this regime it is more appropriate to compare the
results with mean field theory  The mean field results for the same parameters
are also given in figure 6 (dashed line) and  can be seen to be  in good
agreement. Also shown is the Wilson ratio $R(h)/4$, which is only weakly
enhanced at $h=0$, $R(0)=1.244$, and makes a slow crossover to that for
non-interacting electrons for $h\sim \Delta$.\par  
\begin{figure}
\begin{center}
\vskip.4cm
\includegraphics[width=0.42\textwidth,height=5cm]{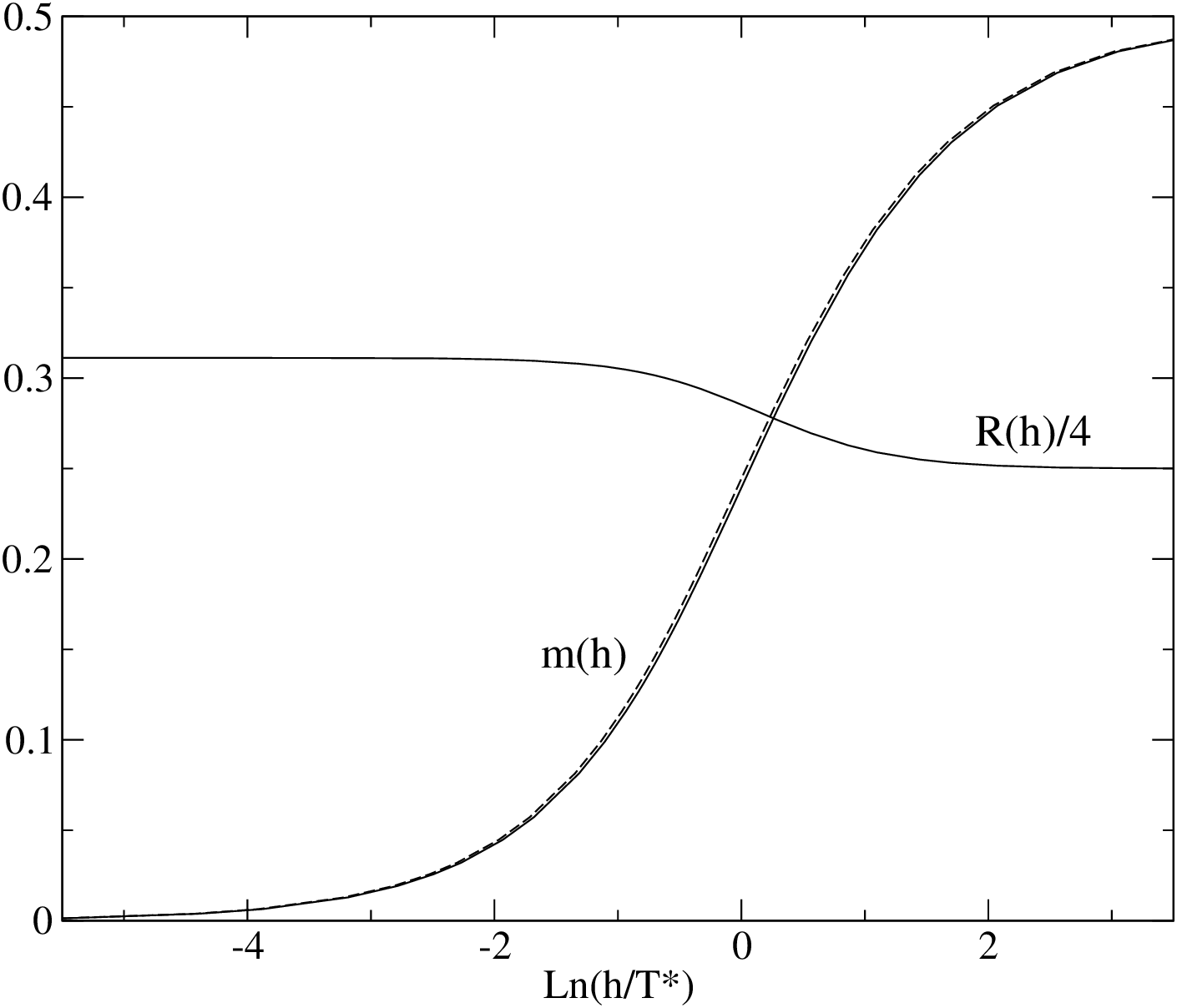}
\smallskip
\label{figure6}
\end{center}
\caption{The impurity magnetization $m(h)$  for the symmetric model for a weak
correlation case with 
  $U /\pi\Delta=0.25$, together with $R(h)/4$, where $R(h)$ is the
 Wilson ratio, plotted as a function of the logarithm of the
 magnetic field. Also shown for comparison are the corresponding results
 (dashed line) for
 the magnetization calculated
using mean field theory.}
\end{figure}

In figure 7 we give the magnetization $m(h)$ against $h/T^*$ for the more
realistically realizable magnetic field regime $0<h<2.5T^*$ for a strong
correlation case  $U/\pi\Delta=3$, using 
 the renormalized parameters in equation (\ref{mag}). Points ($\times$)
 corresponding to Bethe ansatz results \cite{TW83} for the Kondo model, are
 included for comparison and can be seen to be in complete agreement with
 those for the Anderson model in this regime. The results for $m(h)$ for the weak coupling case
$U/\pi\Delta=0.25$  are also shown in figure 7 together with the corresponding
mean field theory results (dashed line). The 
$T^*$s for the strong and weak correlation cases are very different,
$T^*(3)/T^*(0.25)=0.0827$, so the energy scales are very different but, relative to
these scales, $m(h)$ increases initially more slowly with $h$ in the weak
correlation case but approaches saturation more rapidly.
The mean field results  $U/\pi\Delta=0.25$ also give a good approximation to $m(h)$
for $U/\pi\Delta=0.25$ over  this range.\par 

\begin{figure}
\begin{center}
\vskip0.4cm
\includegraphics[width=0.42\textwidth,height=5cm]{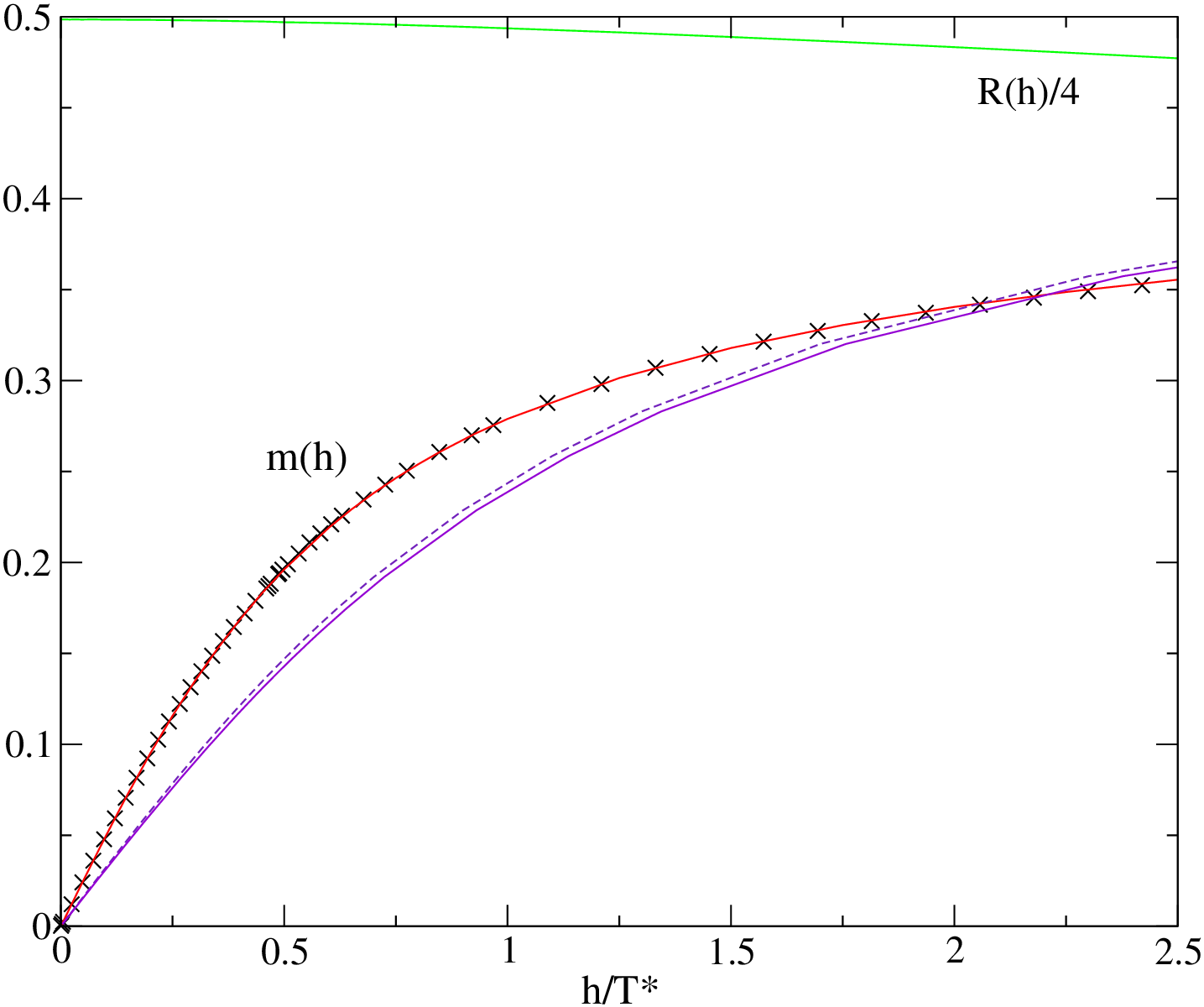}
\smallskip
\label{figure7}
\end{center}
\caption{The impurity magnetization $m(h)$ and $R(h)/4$, where $R(h)$ is the
 Wilson ratio, for the symmetric model with  $U /\pi\Delta=3.0$
  plotted as a function  of the
 magnetic field $h/T^*$, and compared with the Bethe ansatz results \cite{TW83} (crosses)
 for the Kondo model. Also shown are the results for $m(h)$ calculated for
the weak coupling case $U/\pi\Delta=0.25$ together with the corresponding mean
 field results (dashed line).}
\end{figure}

Also shown in figure 7 are  the  values of $R(h)/4$ for the strong
correlation case $U/\pi\Delta=3$.
It can be seen that there is a $5\%$ reduction in $R(h)$ over this range to
$h=2.5T_{\rm K}$, indicating some evidence of charge fluctuations beginning
to contribute to the specific heat coefficient $\gamma_{\mathrm{d}}(h)$ for
intermediate field values, but having  little effect on the
magnetization. The localized model gives $R(h)=2$ for all  $h$,  implies that $\tilde
 U(h)=1/\tilde\rho_{\mathrm{d}}(0,h)$. From this result, and equations
 (\ref{hqpdos}) and (\ref{mag}), the ratio $\tilde 
 U(h)/\pi\Delta(h)$ for the localized model can be expressed entirely in terms
 of the magnetization and is such that   $\tilde
 U(h)/\pi\tilde\Delta(h)=1/\cos^2(\pi m(h))$.
 For $h=0$, this corresponds to
 the strong correlation results  $\tilde
 U(0)/\pi\tilde\Delta(0)=1$, as $m(0)=0$, and for very large fields  where
 $m(h)\to 1/2$ as $h\to
 \infty$, it gives  $\tilde
 U(h)/\pi\tilde\Delta(h)\to\infty$, corresponding to the fact that charge
 fluctuations can only be  completely   suppressed if $U$ is infinite. In
 figure 8 we plot the ratio  $\tilde
 U(h)/\pi\tilde\Delta(h)$ as a function of ${\rm ln}(h/T^*)$ for
 $U/\pi\Delta=3,5$, and compare the results with  $1/\cos^2(\pi m(h))$,
where $m(h)$ is calculated from the Bethe ansatz results for the Kondo
 model. It can be seen that the large peak in  $\tilde
 U(h)/\pi\tilde\Delta(h)$ develops as $h$ increases, because it  closely follows the result
 for  the localized model  $1/\cos^2(\pi m(h))$, and then falls back in very
 large fields to the unrenormalized ratio $U/\pi\Delta$. The results for
 $U/\pi\Delta=5$ follow the form for the localized model  $1/\cos^2(\pi m(h))$,
  implying $R(h)=2$,  for a
 greater range of $h$  than for the case
 $U/\pi\Delta=3$, resulting in a more pronounced peak.
\begin{figure}
  \begin{center}
\vskip0.4cm
    \includegraphics[width=0.4\textwidth,height=5cm]{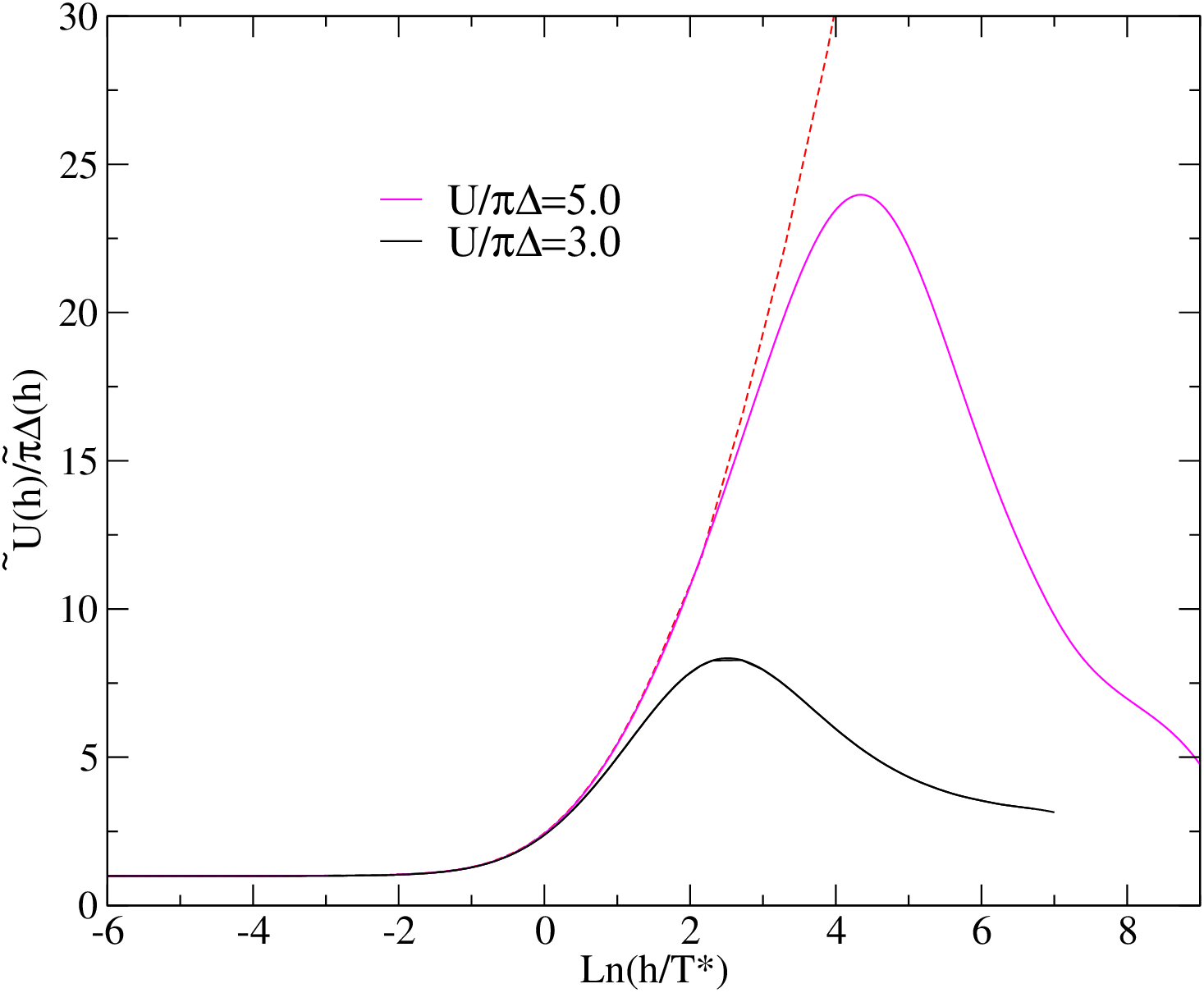}
    \smallskip
    \label{figure8}
  \end{center}
  \caption{The ratio   $\tilde
    U(h)/\pi\tilde\Delta(h)$ is plotted as a function of ${\rm ln} (h/T^*)$
      for the cases
      $U/\pi\Delta=3, 5$. The two results are  compared with the result $1/\cos^2(\pi
      m(h))$, where $m(h)$ is calculated from the Bethe ansatz results \cite{TW83}
      for  the localized model where all impurity charge fluctuations are suppressed.}
\end{figure}

\section{Low Temperature Thermodynamics in an Arbitrary Magnetic Field }
We can generalize some of the results in the preceding section to calculate
the leading temperature dependence in the presence of an arbitrary magnetic
field. For instance, for the temperature dependence of the susceptibility and
magnetization
we can use the thermodynamic relation,
\begin{equation} {\partial^2C(T,H)\over\partial H^2}=T
{\partial^2\chi(T,H)\over\partial T^2},\end{equation}
where $C(T,H)$ is the specific heat. Applying this to the impurity
contribution and taking the limit $T\to 0$ we find
\begin{equation}\left. {\partial^2\chi_s(T,H)\over\partial T^2}\right|_{T=0}=
{\partial^2\gamma_{\mathrm{d}}(H)\over\partial H^2}.\label{tdr}\end{equation}
We can use the result in equation (\ref{rgam}) to deduce the results for the
$T^2$ dependence for $\chi_s(T,h)$,
\begin{eqnarray*}
{\chi_s(0,h)-\chi_s(T,h)\over \chi_s(0,h)}&=&-{\pi^2\over 12} {T^2\over
  \chi_s(0,h)}  {\partial^2\tilde\rho_{\mathrm{d}}(0,h)\over\partial
  h^2} \\
&=&c_{\chi}(h)\left(T\over T^*\right)^2.
\end{eqnarray*}
On integrating these results with respect to $h$ we can derive a similar
relation for the induced magnetization, $M(T,h)=m(T,h)/(g\mu_{\rm B})$,
\begin{equation}
{m(0,h)-m(T,h)}=-{\pi^2\over 6} {T^2}
  {\partial\tilde\rho_{\mathrm{d}}(0,h)\over\partial h}=c_{m}(h)\left(T\over
  T^*\right)^2\end{equation}
We can calculate the coefficients  $c_{\chi}(h)$ and $c_{m}(h)$ analytically
  in the non-interacting case, 
\begin{equation}
c^{(0)}_{\chi}(h)={\pi^4\over 48}{\left({1-3\left({\pi h/
        4T^*}\right)^2}\right)\over 
\left[1+\left({\pi h/4T^*}\right)^2\right]^2},\label{cchi}
\end{equation}
\begin{equation}c^{(0)}_{m}(h)={\pi^4\over 48}{\left({ h/ 4T^*}\right)\over
\left[1+\left({\pi h/ 4T^*}\right)^2\right]^2},
\end{equation}
where $T^*=\pi\Delta/4$.\par
We can also deduce the values of $c_{\chi}(h)$ and $c_{m}(h)$ 
in the Kondo regime from the Bethe ansatz results for $\chi_{\mathrm{d}}(0,H)$.
As the Wilson or '$\chi/\gamma$' ratio $R$ has the value 2 in this regime,
independent $H$  as the charge 
fluctuations can be neglected, the value of $\gamma_{\mathrm{d}}(H)$ can be deduced
and is  proportional to
$\chi_{\mathrm{d}}(0,H)$. On substituting the value for $\gamma_{\mathrm{d}}(H)$ into
equation (\ref{tdr})  $c_{\chi}(h)$ and $c_{m}(h)$ can then be calculated.
The asymptotic values as $h\to 0$ are  $c_{\chi}(0)=\sqrt{3}\pi^3/8+{\rm
  O}(h^2)$ and $c_{m}(h)=h\sqrt{3}\pi^3/16T_{\rm K}+{\rm O}(h^3)$.\par 
\begin{figure}
  \begin{center}
\vskip.4cm
    \includegraphics[width=0.4\textwidth,height=5cm]{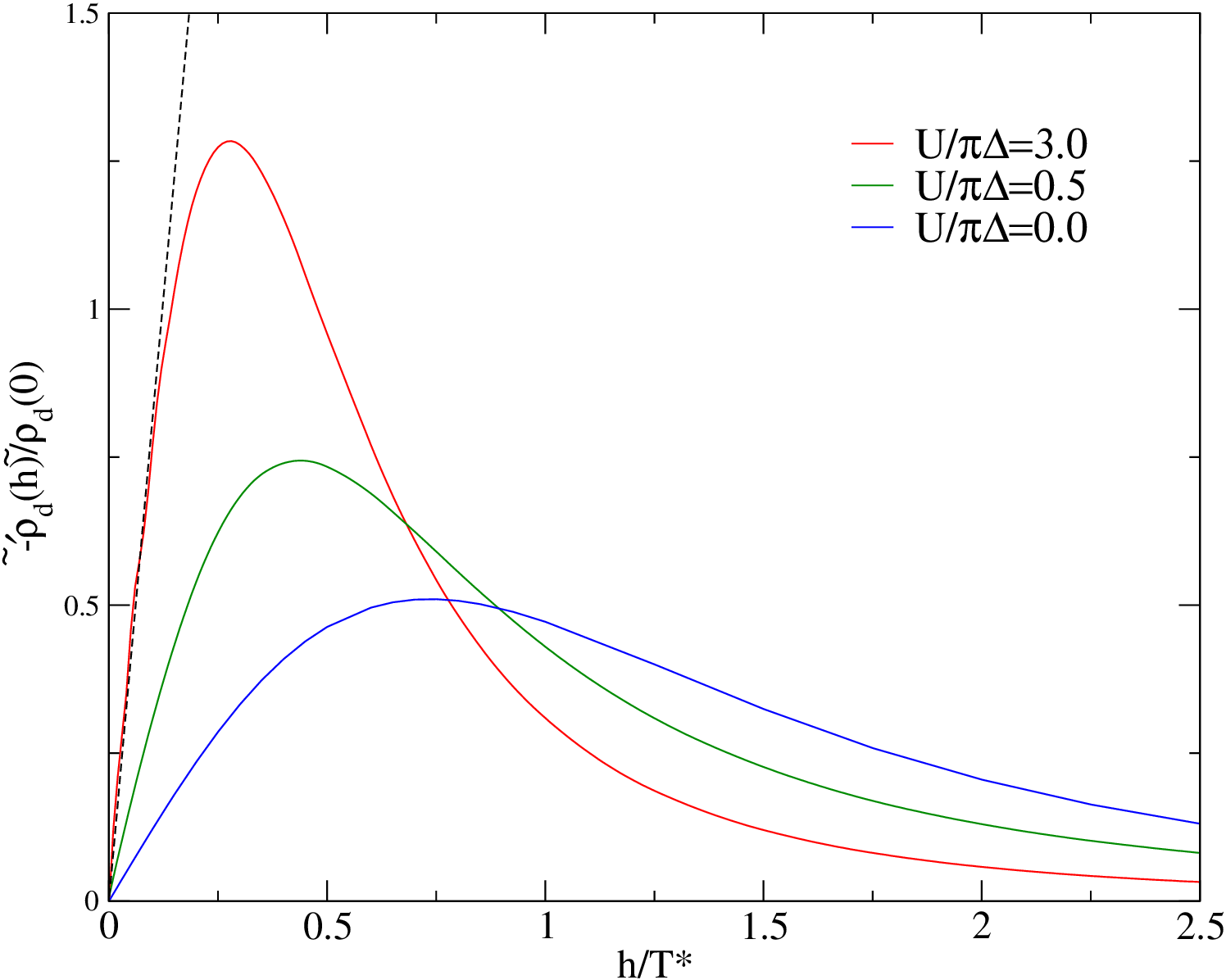}
    \smallskip
  \end{center}
  \caption{The ratio
    $-\tilde\rho'_{\mathrm{d}}(h)/\tilde\rho_{\mathrm{d}}(0)$, where the prime 
    indicates a derivative with respect to $h/T^*$, is shown for
    $U/\pi\Delta=3.0,0.5,0.0$  as a function of $h/T^*$. The dotted line shows
    the asymptotic result as $h\to 0$, $3h\pi\sqrt{3}/2T^2_{\rm K}$, for the
    Kondo model.  } 
    \label{figure9}
\end{figure}

In figure 9 we plot the results for $-\tilde\rho'_{\mathrm{d}}(h)/\tilde\rho_{\mathrm{d}}(0)$,
which is proportional to $c_m(h)$, for  $U/\pi\Delta=3.0,0.5,0.0$
in the range $0< h/T^*<2.5$.
It can be seen that all three curves  have a maximum which
implies that $c_{\chi}(h)$ has a zero, and therefore changes sign from
positive to negative in this range. In the non-interacting case it can be seen
from the result in equation (\ref{cchi})     
that the change of sign occurs when $h/T^*=4/\pi\sqrt{3}$, and from figure \ref{figure9},
that it occurs for $h$ significantly smaller than $T_{\rm K}$ in the Kondo
regime. 
  
\section{Low Temperature Transport in an Arbitrary Magnetic Field }

To extend the calculations to the low energy dynamics we use the renormalized
perturbation theory\cite{Hew93,Hew01}, which is  described briefly in the Appendix.
 In this form of perturbation theory we work with the renormalized parameters,
$\tilde\epsilon_{\mathrm{d}}$, $\tilde\Delta$  and $\tilde U$ instead of the
original bare parameters $\epsilon_{\mathrm{d}}$, $\Delta$ and $ U$. The
free propagators are those of the non-interacting quasiparticles and the
expansion is in 
powers of $\tilde U$. As the parameters are already fully renormalized counter
terms have to be introduced to cancel any further renormalization.
This expansion is completely
specified by the three parameters $\tilde\epsilon_{\mathrm{d}}$,
$\tilde\Delta$  and $\tilde U$, and is not restricted to
the low energy regime, but valid for all energy scales. In practice,
however, the calculations are easier to carry out in the low energy regime, where
asymptotically exact results can be obtained in this regime by working only to
second order in $\tilde U$.\par
Here we exploit the fact that we have these renormalized parameters as a
function of arbitrary magnetic field for the symmetric model to calculate the
low energy dynamics  in the presence of a magnetic field.
The quasiparticle retarded  Green 
function for the impurity level $\tilde G_{\mathrm{d}}(\omega,T,h)$ takes the form
\begin{equation}
\tilde G_{\mathrm{d},\sigma}(\omega,T,h)=
  \frac{1}{\omega+\sigma\tilde\epsilon_{\mathrm{d}}(h)+
  i\tilde\Delta(h)-\tilde\Sigma_\sigma(\omega,T,h)}, 
\label{qpgfct}
\end{equation}
where $\tilde\Sigma_\sigma(\omega,T,h)$ is the renormalized self-energy.

To calculate the leading order $T^2$ term in the transport
coefficients we need  $\tilde\Sigma_\sigma(\omega,T)$     both to order $\omega^2$  
and to order $T^2$. We calculate this  from the
renormalized perturbation expansion taken to order $\tilde U^2(h)$.  This
takes full  account of the quasiparticle scattering and gives the exact result of
Yamada\cite{Yam75b} for $h=0$.  
The corrections to order $\omega^2$ can be deduced
from the second derivative of the self-energy with
respect to $\omega$ evaluated at $\omega=0$ and $T=0$. The result for the
renormalized self-energy to this order is given by
\begin{equation}
\tilde\Sigma_{\sigma}(\omega,0,h)=
-\tilde\alpha(h)\omega^2\left[i
-(2+ \tilde\alpha_\omega(h)) { \sigma\tilde\varepsilon(h)}\right],
\end{equation}
where $\tilde\varepsilon(h)=\tilde\epsilon_{\mathrm{d}}(h)/\tilde\Delta(h)$,
and $\tilde\alpha(h)$ is given by  
\begin{equation}
\tilde\alpha(h)=\frac{\pi}2\tilde\rho_{\mathrm{d}}(0,h)(R(h)-1)^2,
\end{equation}
and  $\alpha_\omega(h)$ by
\begin{equation}
\tilde\alpha_\omega(h)={2I(h)\tilde\Delta(h)\over\tilde\xi(h)\tilde\rho^2_{\mathrm{d}}(0,h)},
\end{equation} 
where $\tilde\xi(h)=\pi\tilde\rho_{\mathrm{d}}(0,h)\tilde\Delta(h)\tilde\varepsilon(h)$
and $I(h)$ is the integral
\begin{equation}
I(h)=\int_{-\infty}^{\infty} 
\int_{-\infty}^{\infty}
\tilde G^0_{\downarrow}(\omega'')\tilde G^0_{\downarrow}(\omega''+\omega')
[\tilde G^0_{\uparrow}(\omega')]^3{d\omega''\over 
  2\pi}{d\omega'\over 2\pi}. 
\end{equation}  
$\tilde G^0_{\sigma}(\omega)$ is the propagator in the diagrammatic
 expansion for $T=0$ which is given by
\begin{equation}
 [\tilde G^0_{\sigma}(\omega)]^{-1}=
\omega+\sigma\tilde\epsilon_{\mathrm{d}}(h)+{\rm sgn}(\omega)i\tilde\Delta(h).
\end{equation}
The corresponding result for the renormalized self-energy to order $T^2$ can
be derived using the Sommerfeld 
expansion. The calculation can be performed by using for each internal
 propagator  $\tilde G^0_{\sigma}(\omega)$
 in   the  $T=0$ diagrammatic expansion an additional correction term, 
\begin{equation}
-{(\pi T)^2\over 3}
\frac{\delta'(\omega)\tilde\Delta(h)}
{(\omega+\sigma\tilde\epsilon_{\mathrm{d}}(h))^2+\tilde\Delta^2(h)}.
\end{equation}
The result for the renormalized self-energy to order $T^2$ for $\omega=0$ is
\begin{equation}
\tilde\Sigma_\sigma(T,0,h)=-\tilde\alpha(h)
    (\pi T)^2\left[i+\sigma\tilde\varepsilon(h)(1+\tilde\alpha_T(h))\right],
\label{seT}
\end{equation}
where the parameter $\tilde\alpha_T(h)$ is given by
\begin{equation}
\tilde\alpha_T(h)={1\over
    6\tilde\xi(h)\tilde\varepsilon(h)}\left[1-\tilde\varepsilon(h){\rm
    tan}^{-1}(\tilde\varepsilon(h))\left(
4+{1\over\tilde\xi(h)\tilde\varepsilon(h)}\right)\right].
\end{equation}
We can
now apply these results to the calculation of transport coefficients.

\subsection{Application to magnetic impurities}
The contribution to the conductivity from the scattering of isolated
impurities described by an AIM, $\sigma(T,h)$, given by\cite{Yam75b}
\begin{equation} 
\sigma(T,h)= \sigma_0\sum_\sigma \int_{-\infty}^{\infty} 
   \frac{1}{\rho_{d,\sigma}(\omega,T,h)}\left (-{\partial
   f(\omega)\over\partial\omega}\right) d\omega, 
\label{int}  
\end{equation}
where $\rho_{\mathrm{d}}(\omega,T,h)=\tilde\Delta(h)\tilde\rho_{\mathrm{d}}(\omega,T,h)/\Delta$,
and $\tilde\rho_{\mathrm{d}}(\omega,T,h)$ is the spectral density of the quasiparticle 
Green function $\tilde G_{\mathrm{d}}(\omega,T,h)$. The Sommerfeld expansion
gives for (\ref{int}) to second order in $T$,
\begin{eqnarray}
&&  \sigma(T,h)=
\sigma_0\Big\{\sum_{\sigma}\frac1{\rho_{\mathrm{d},\sigma}(0,T,h)}+\label{somexp} \\
&&  \Big[\sum_{\sigma}\Big(\frac{\rho_{d,\sigma}'(0,0,h)^2}{\rho_{d,\sigma}(0,0,h)^2}-
  \frac{\rho_{d,\sigma}''(0,0,h)}{\rho_{d,\sigma}(0,0,h)}\Big)\Big]\frac{(\pi
  T)^2}{6}\Big\}, \nonumber
\end{eqnarray}
where the prime refers to a derivative with respect to $\omega$. Note that
the first term still contains a temperature dependence via
$\rho_{\mathrm{d}}$.

On using the renormalized self-energy to calculate the quasiparticle spectral
density  $\tilde\rho(\omega,T,h)$,
and substituting in equation (\ref{somexp}), the final result for $\sigma(T,h)$
to order $T^2$  is\cite{comment1}
\begin{equation}
\sigma(h,T)=\sigma(h,0)
\Big\{1+\sigma_2(h)\left({\pi T\over \tilde\Delta(h)}\right)^2+{\rm 
    O}(T^4)\Big\},
\end{equation} 
where $\sigma(h,0)=2\sigma_0/ {\rm cos}^2(\pi m(h))$ and $\sigma_2(h)$ is given by
\begin{equation}
\sigma_2(h)={{\rm cos}^2\pi
    m(h)\over 3}\left[1+C(h)(R(h)-1)^2 \right].
\label{s2}
\end{equation}
The coefficient $C(h)$ is
 \begin{equation}
C(h)=2{\rm cos}^2(\pi m(h))-{\rm sin}^2(\pi
m(h))\left[1-3\tilde\alpha_T(h)+\tilde\alpha_\omega(h)\right].
\end{equation}

\begin{figure}[htb]
  \centering
  \includegraphics[width=0.45\textwidth,height=5cm]{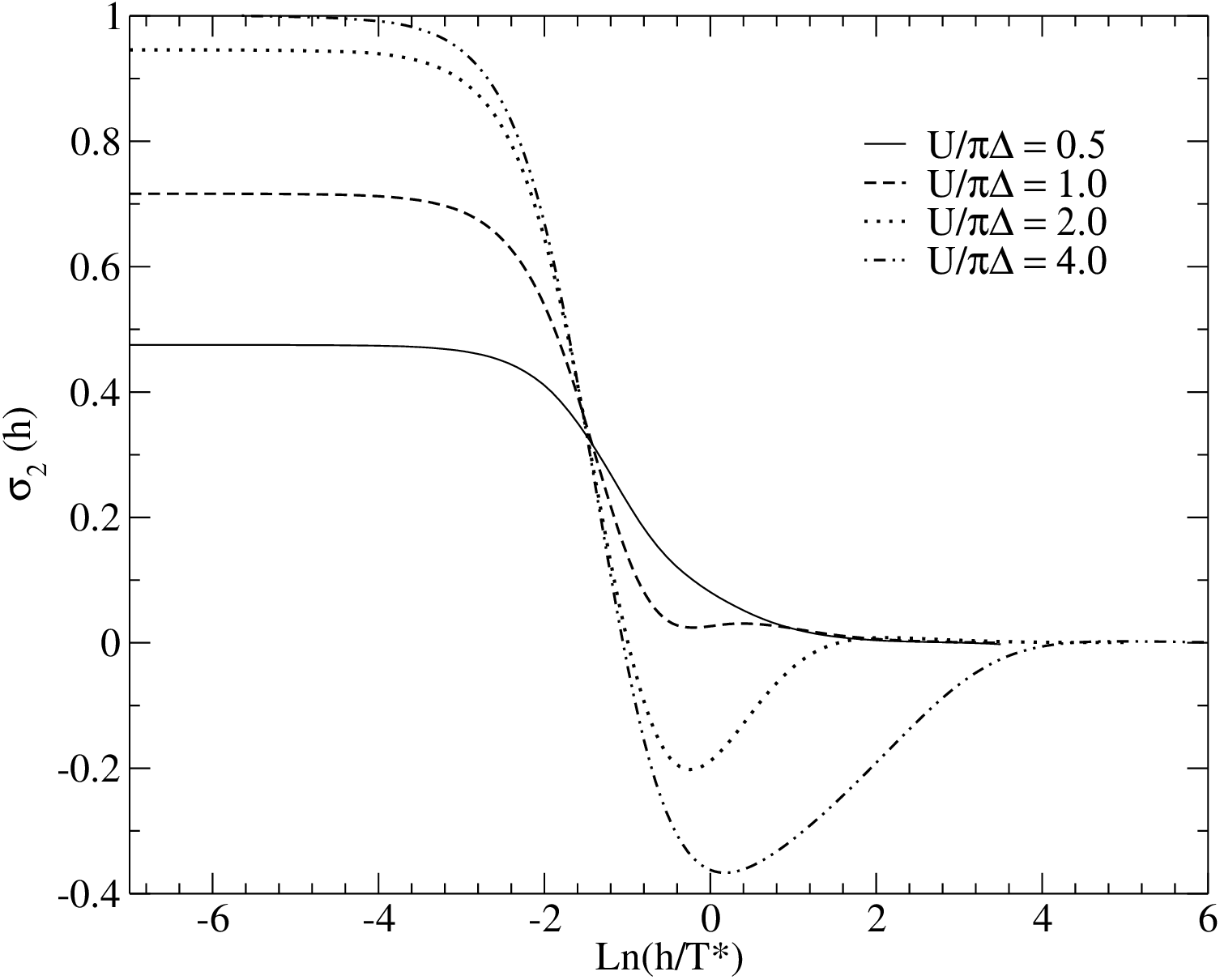}\\[0.5cm]
  \includegraphics[width=0.45\textwidth,height=5cm]{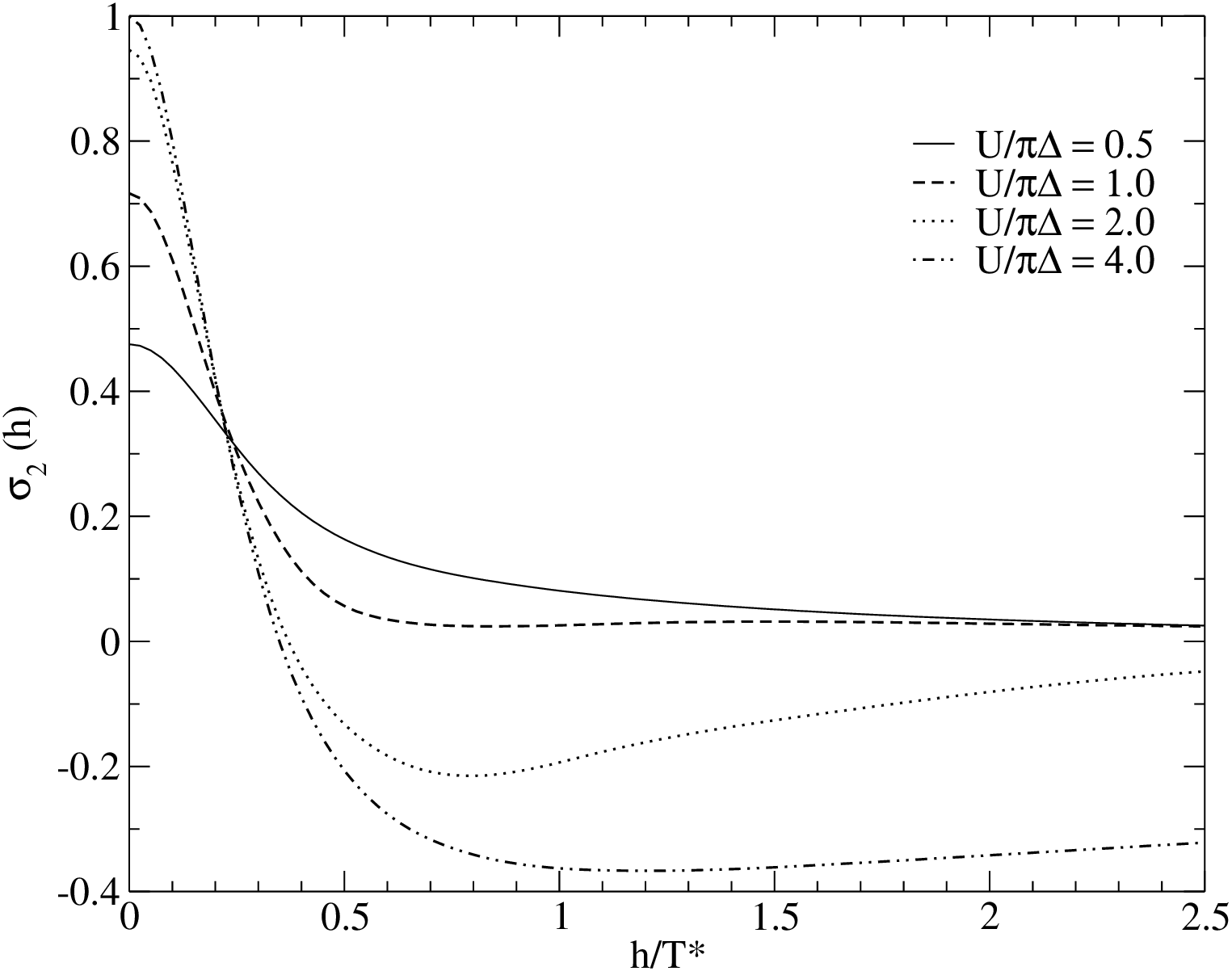}
  \caption{Field dependent coefficient $\sigma_2(h)$ from (\ref{s2}) for the second order
    temperature expansion of the conductivity. Weak coupling ($U/\pi\Delta=0.5$)
    up to strong coupling ($U/\pi\Delta=4$) is considered.} 
  \label{sig2}
\end{figure}

\noindent

In figure \ref{sig2} we show the second order coefficient $\sigma_2(h)$
plotted over $\log(h/T^*)$ for a range of parameters
($U/\pi\Delta=0.5-4$). For zero field the conductivity rises with  
temperature as is well known \cite{Yam75b}.   
When $h$ is increased $\sigma_2(h)$ decreases and tends to zero for very high
fields, so that the low temperature conductivity becomes temperature independent.
The impurity level is then shifted out of the range of the thermally excited
states in the conduction band so that there is negligible impurity scattering.  
We note for the strong coupling cases where there is a local moment ($U/\pi\Delta=2,4$), the coefficient
$\sigma_2(h)$  changes sign for a certain critical field $h_c$, with $h_c\sim
0.5T^*$. This does not arise, as one could expect from the change of the
impurity density of states at the Fermi level with the field [second term in  
(\ref{somexp})], but from the temperature dependence of the self-energy, which
contributes in the first term in (\ref{somexp}). To see this only consider the
imaginary part contribution to the self-energy, \ref{seT}, which leads to a $T^2$
term  
\begin{equation}
(\pi T)^2/2(R(h)-1)^2[\cos^2(\pi m(h))-\sin^2(\pi m(h))].
\end{equation}
This expression clearly changes sign when $m(h)$ exceeds $1/4$ or
  $\tilde\epsilon_{\rm  d}(h)>\tilde \Delta(h)$. In contrast the contribution from
  the second term in  (\ref{somexp})  with only the imaginary part of the
  self-energy taken into account gives 
\begin{equation}
(\pi T)^2/3(1+(R(h)-1)^2[\cos^2(\pi m(h))-\sin^2(\pi m(h))]),
\end{equation}
which cannot change sign. Including the real part
contribution does not alter this behaviour substantially.

Physically, it is the spin flip scattering of the local moment that causes the
resistance to rise as the temperature is lowered, leading to
a resistance minimum and the Kondo effect. Perturbation theory shows that
spin-flip scattering gives a diverging amplitude for $T\simeq T^*$. For a mainly
polarized impurity spin these processes are, however, strongly suppressed, and 
the change in sign of the temperature dependence might be attributed to the
thermal spin disorder scattering.
To our knowledge, this effect  has not been observed   but for magnetic 
 impurities systems with a very low  Kondo temperature it might be feasible to
put the result to an experimental test. 

\subsection{Application to quantum dots}
The Kondo effect in a magnetic field has been observed experimentally  
in mesoscopic systems, for example, 
quantum dots in heterostructures\cite{Kas92}. Such systems 
can be quantitatively described by  the Anderson impurity
model\cite{PG04,GKSMM98}. The symmetric Anderson model corresponds to the
situation, where the gate voltage $V_g$ is tuned to the  
middle of a Coulomb valley with odd number of electrons on the dot.

Hershfield et al. \cite{HDW91} and Meir and Wingreen \cite{MWL93} derived an
expression for the current through a quantum 
dot by a non-equilibrium calculation, which for the case of symmetric coupling to
the leads takes the form
\begin{eqnarray}
 I&=&
 \frac{G_0\Delta}{2e}
 \sum_{\sigma}\int d\omega(-\Imag
 G^{\mathrm{noneq}}_{\mathrm{d}\sigma}(\omega))
 \times \nonumber \\
&&\times [n_{\rm F}(\omega-\mu_L)-n_{\rm F}(\omega-\mu_R)],
\label{current2}
\end{eqnarray}
where $n_{\rm F}$ is the Fermi function, $G^{\mathrm{noneq}}_{\mathrm{d}s}$ is the retarded
non-equilibrium Green function from the Keldysh formalism on the dot,
$\mu_L=\mu+eV/2$, $\mu_R=\mu-eV/2$ are the chemical potentials in left, right
lead, respectively, and $V=V_{\rm ds}$ the source drain voltage.
\cite{comment2}
$G_0={e^2}/{\pi\hbar}$ with Planck's constant $\hbar$.

In the limit of linear response the equilibrium value of the 
one-electron Green function can be used to evaluate (\ref{current2}),
 and the resulting expression for the differential conductance
$G=dI/dV$ through a quantum dot is
\begin{equation}
  G(T,h)=\frac{G_0\Delta}2\sum_{\sigma}\int d\omega{}{}
  \pi\rho_{\mathrm{d},\sigma}(\omega,T,h)
  \left(-\partdera{\omega}{n_{\rm F}(\omega)}\right).
\label{Gtemp}
\end{equation}
In the low  temperature regime we can again apply the Sommerfeld expansion
to obtain the leading order finite temperature corrections to order $T^2$, 
\begin{eqnarray}
  G(T,h)&=&\frac{G_0\Delta}2\Big[\sum_{\sigma}\pi\rho_{\mathrm{d},\sigma}(0,T,h)+ \nonumber\\
&& \Big(\sum_{\sigma}\pi\rho_{\mathrm{d},\sigma}''(0,0,h)\Big) 
 \frac{(\pi T)^2}{6}\Big].
\label{gth}
\end{eqnarray}
Here, in contrast to the earlier case the second term changes
sign for a critical field $h_c$ when
$\sum_{\sigma}\rho_{\mathrm{d},\sigma}(\omega,0,h)$  changes from a maximum to a
local minimum at $\omega=0$. For free
quasiparticles this happens when $\tilde\epsilon_{\rm  d}(h_c)=\tilde
\Delta(h_c)/\sqrt{3}$, as discussed in detail in reference\cite{HBO05pre}.
The temperature dependent part of the first term in equation (\ref{gth})
including only the imaginary part from the self-energy (\ref{seT}) is
\begin{equation}
  \frac1{2\tilde \Delta}(R(h)-1)^2\frac{\sin^2(\pi m(h))-\cos^2(\pi
  m(h))}{(1+\tan^2(\pi m(h)))^2},
\end{equation}
and this also changes sign for $m(h)>1/4$.
The total result is 
\begin{equation}
  G(T,h)=G(0,h)\left(1-G_2(h)\left(\frac{\pi T}{\tilde\Delta(h)}\right)^2
  \right),
\label{Gtemp2}
\end{equation}
with
\begin{equation}
  G(0,h)=G_0\cos^2(\pi m(h)),
\end{equation}
and
\begin{eqnarray*}
  G_2(h)=\frac{\cos^2(\pi m(h))}{3}\Big\{\!\!\cos^2(\pi m(h))\left[1+2(R(h)-1)^2\right] \\
 -\sin^2(\pi
m(h))\!\!\left[3+(R(h)-1)^2(1+2\alpha_{\omega}(h)-6\alpha_T(h))\right]\!\!\Big\}.
\label{g2}
\end{eqnarray*}

\begin{figure}[htb]
  \centering
  \includegraphics[width=0.42\textwidth,height=5cm]{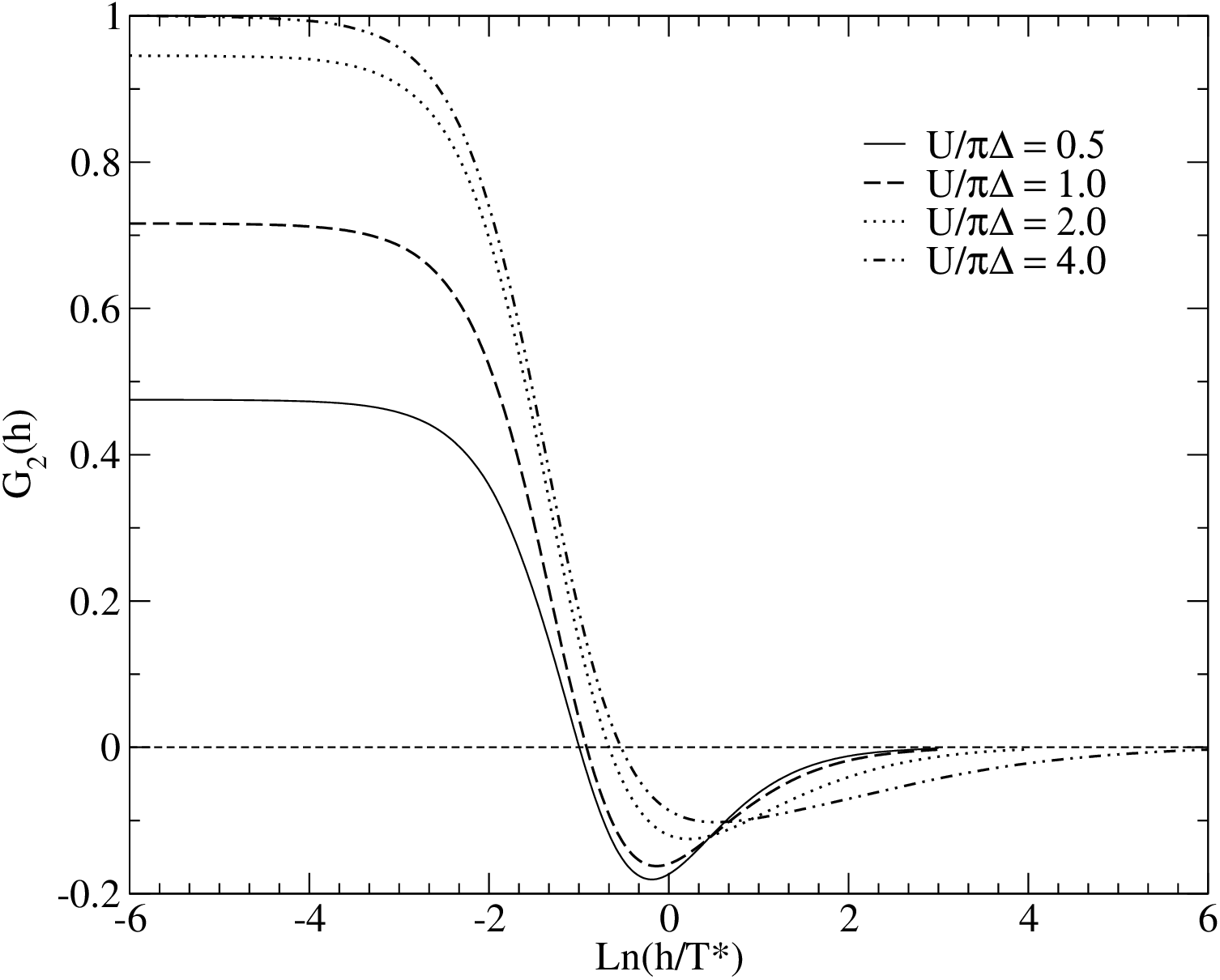}\\[0.5cm]
  \includegraphics[width=0.42\textwidth,height=5cm]{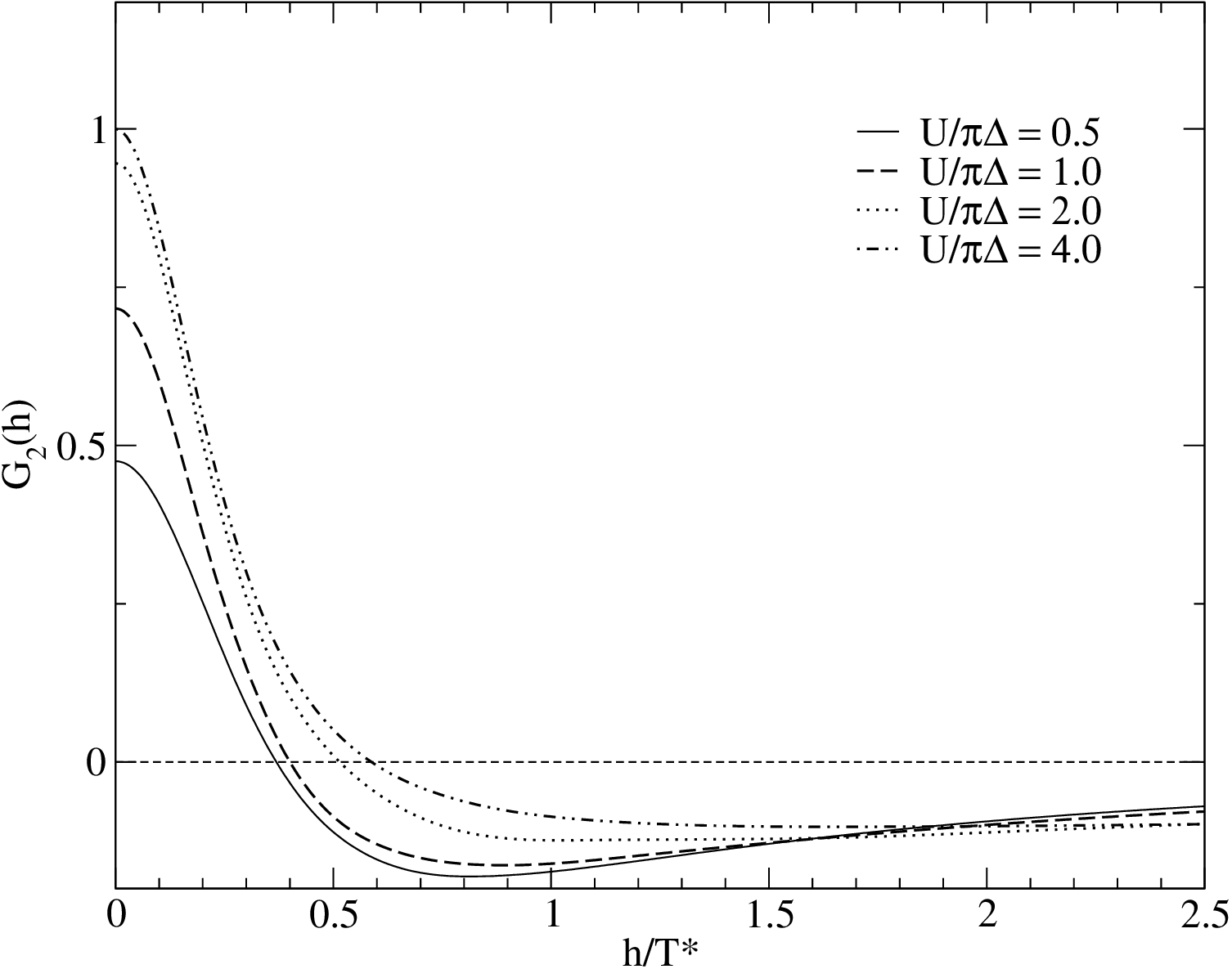}
  \caption{Field dependent coefficient $G_2(h)$ for the second order
    temperature expansion of the conductance. Weak coupling ($U/\pi\Delta=0.5$)
    up to strong coupling ($U/\pi\Delta=4$) is considered.} 
  \label{g2fig}
\end{figure}

\noindent
In figure \ref{g2fig} the field dependence of $G_2(h)$ is shown.
Note that we have included a minus sign before the $T^2$ term in
(\ref{Gtemp2}), so that the similar behaviour in figures \ref{sig2} and
\ref{g2fig} actually corresponds to opposite temperature dependence.
This is due to the approximate inverse relation between the two systems,
if the hybridization $V_{k}=0$ for an impurity, there is no scattering
and hence infinite conductivity, whereas if $V_k=0$
for the quantum dot there is no current and hence infinite resistivity. \\  

\section{Beyond the Low Energy Regime}

\begin{figure}[!htb]
\begin{center}
\psfrag{w}{$\omega$}
\psfrag{rho_u}{$\rho_{\mathrm{d},\uparrow}(\omega)$}
\includegraphics[width=0.42\textwidth,height=6cm]{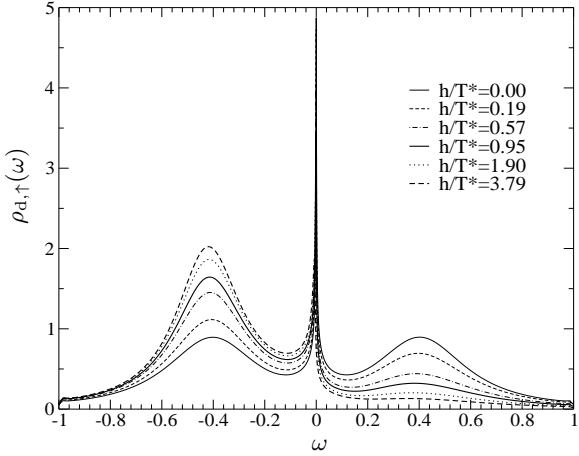}
\vskip-.2cm
\end{center} 
\caption{Strong coupling ($U/{\pi \Delta}=4$) spectral density of the d-site Green function
  $\rho_{\mathrm{d},\uparrow}(\omega)$ for various magnetic fields $h$. The
  energy scale is given by $4T^*=\pi\tilde\Delta$.}
\label{spectra}
\end{figure}

We can use the extension of the NRG method to the calculation of dynamic
response functions\cite{SSK89,CHZ94} to look at the behaviour of the model in an
arbitrary magnetic field on higher energy scales.
In doing so it is important to use  the density matrix (DM-NRG)
method introduced by Hofstetter\cite{Hof00} as the standard NRG approach gives
results which considerably underestimate the shift of the high energy spectral
weight with  the variation of magnetic field. We also use the approach of
Bulla et al.\cite{BHP98}, in which the self-energy is deduced from the
calculation of higher order Green functions, as this gives more accurate
results.
In figure \ref{spectra} we give results for the spin up part of the d-site spectral density $\rho_{\mathrm{d},\uparrow}(\omega)=-\frac
1{\pi}\Imag G_{\mathrm{d},\uparrow}(\omega^+)$ for a strong coupling situation ($
U/\pi \Delta=4$) for various values of the magnetic field $h$.
The shift of the spin-up Kondo resonance from the Fermi level with increase of
magnetic field, which is almost imperceptible
on the scale shown, is accompanied by large shifts of the spectral weight on the
higher energy scales as the impurity is magnetically polarized. \\

\begin{figure}[!htb]
\begin{center}
\psfrag{ln(h/pidel)}{${\rm Ln}(h/(\pi\Delta))$}
\psfrag{m(h)}{$m(h)$}
\includegraphics[width=0.42\textwidth,height=6cm]{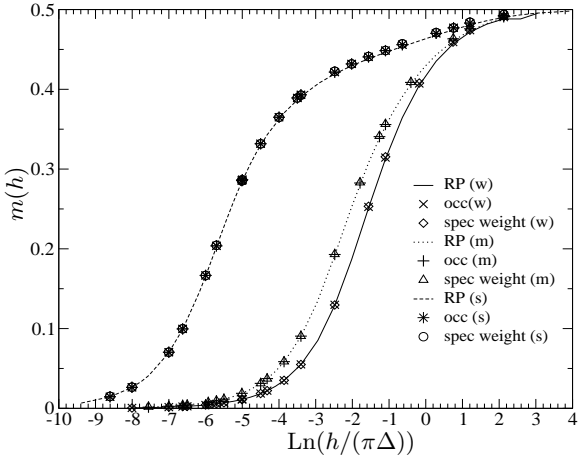}
\vskip-0.2cm
\end{center} 
\caption{Comparison of d-site magnetization $m(h)$ for the full range of $h$ for 
  $U/{\pi \Delta}=0.5$, weak coupling (w), $U/{\pi \Delta}=1$,
  intermediate coupling (m) and  $U/{\pi \Delta}=4$, strong coupling
  (s) computed from the renormalized parameters (RP), the occupancy (occ) and
  the spectral weight as described in the text.} 
\label{magn}
\end{figure}

\noindent
In order to test the improved estimates of this shift from the DM-NRG, we calculate
$\langle n_{\uparrow}\rangle$ by integrating the spectral density up to
the Fermi level to deduce the value of the induced magnetization $m$.
This result can then be compared with that calculated earlier from the
renormalized parameters using equation (\ref{mag}), and also with the values calculated directly
from the ground state occupation numbers at the impurity site, as deduced from
the matrix elements in the NRG routine.  The results are shown in figure
\ref{magn} for the three parameter  regimes $U/{\pi \Delta}=0.5,1,4$ over
$\ln(h/\pi\Delta)$. 
The magnetization departs from zero already for very low $h/\pi\Delta$ in the
strong coupling case due to the induced lower energy scale $\tilde \Delta$,
whereas for the intermediate and weak coupling cases higher fields are
required to show similar behaviour. For large $h$ all three merge and approach
a saturation value $m_s=\frac 12$, corresponding to complete
polarization of the electron on the impurity site. 
The agreement of the results is excellent in the three parameter regimes for all
$h$, and confirms that  the spectral weight in a broken symmetry situation can be
computed correctly using the DM-NRG.

\begin{figure}[!htb]
\begin{center}
\vskip.4cm
\psfrag{w}{$\omega$}
\psfrag{rho_u}{$\rho_{\mathrm{d},\uparrow}(\omega)$}
\includegraphics[width=0.42\textwidth,height=6cm]{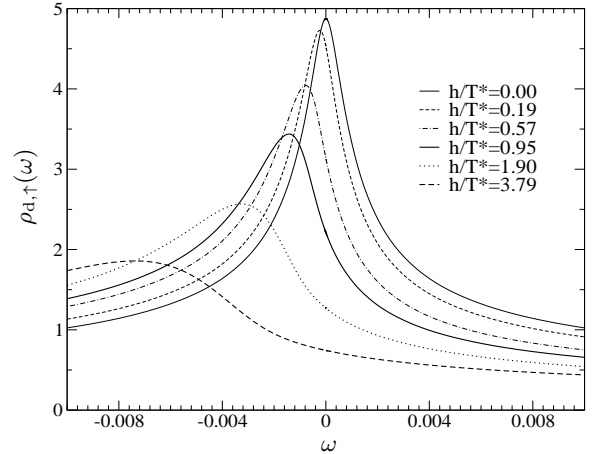}
\vskip-.2cm
\end{center} 
\caption{Quasi particle peak for the spectral density of the d-site Green function
  $\rho_{\mathrm{d},\uparrow}(\omega)$ for various values of $h$ for the same
  parameter set as used for figure \ref{spectra}.} 
\label{spectrazoom}
\end{figure}

\noindent
In figure \ref{spectrazoom}, we focus on the shift of the quasiparticle (Kondo)
resonance in the results for the
strong coupling case ($U/{\pi \Delta}=4$) as in figure \ref{spectra}.  
This shift of the resonance from the Fermi level ($\omega=0$) with increasing
magnetic field values is clearly seen on the higher resolution  energy scale
used for this plot. As the peak shifts, its height decreases and the
resonance becomes broader. For even larger fields than shown here the peak
merges with the lower atomic limit peak seen in figure \ref{spectra}.
Note that the peak form is asymmetric with logarithmic tails, similar to 
the results of Rosch et al. \cite{RCPW03}, obtained using the perturbative RG for the
Kondo model. However, some of the asymmetry in the our results must be
attributed to the logarithmic broadening scheme.

\noindent
If $-\epsilon_p(h)$ is  the position of the quasiparticle peak in the spectral
density for a spin up electron, then the corresponding value for
non-interacting electrons ($U=0$) is half the Zeeman splitting, $\Delta_Z$, where
$\Delta_Z=2h$. An exact expression for  $\epsilon_p(h)/h$ in the limit $h\to
0$ has been derived by   Logan and Dickens\cite{LD01}, 
\begin{equation}
\lim_{h\to 0}{\epsilon_p(h)\over h}=\frac R{1+b \Delta z^2},
\label{smallhform1}
\end{equation} 
where $b$ is the curvature of the imaginary part of the self-energy at $\omega=0$.
The value of $b$ can be calculated from the renormalized perturbation
expansion\cite{Hew01} and the result (\ref{smallhform1}) written as
\begin{equation}
\lim_{h\to 0}{\epsilon_p(h)\over h}=\frac{R}{1+(R-1)^2/2}.
\label{smallhform2}
\end{equation}
This ratio, therefore, varies from 1 in the non-interacting case ($R=1$)
to 4/3 in the Kondo limit ($R=2$).  
It is not straight forward to obtain
a precise estimate of $b$ or the value of $\epsilon_p(h)$ 
from the DM-NRG spectra as they are sensitive to parameters of the logarithmic
scale Gaussian 
broadening which is used to obtain a continuous spectrum on all energy scales 
from the discrete results. However, if the broadening is modified to Lorentzian
peaks with constant width for the very low energy scales
the asymptotic results can be confirmed.

We have estimated the ratio $\epsilon_p(h)/ h$ from the NRG spectra for higher
magnetic field 
values and find that the it increases monotonically with $h$ and 
exceeds the value of 2 before the peak merges at high field values into the
atomic-like peaks.
There have been other estimates of the $h$-dependence of this
ratio\cite{MW00,Cos00,LD01}, but these differ markedly according to the method
of calculation. 
On the basis of a Bethe ansatz calculation of the spinon spectrum for the
Kondo model  Moore and Wen\cite{MW00}   find that  $\epsilon_p(h)/ h<2$ in all cases
and conjecture that the value of $2$ is the high field asymptotic limit. It 
is possible that this is a feature of the localized model, when charge
fluctuations are completely suppressed.
There is some evidence in support of this in our results in that, as we
suppress  the charge fluctuations on increasing the value of $U$ through the
values 
$U/\pi\Delta=2,3,4$, the ratio   $\epsilon_p(h)/ h$   increases less rapidly
with increase of $h$. The ratio only
begin to exceed the value of $2$ roughly at the point when charge fluctuations
set in and $R(h)$ begins to differ significantly from the value of $R(h)$ for
the localized model, $R(h)=2$. Costi\cite{Cos00} has also done
NRG calculations for a localized model and finds a ratio close to but  always less
than 2. Using the local moment approximation Dickens and Logan\cite{LD01} have also 
estimated the ratio $\epsilon_p(h)/ h$ and find an even more marked increase
in the ratio with increase of $h$ to values such that $\epsilon_p(h)/ h>2$. \\

\begin{figure}[!htb]
\begin{center}
\vskip.4cm
\psfrag{tkotk0}{$T_{\rm K}/T_{\mathrm{K},0}$}
\psfrag{epsoh}{${\epsilon_p}/h$}
\centering
\includegraphics[width=0.42\textwidth]{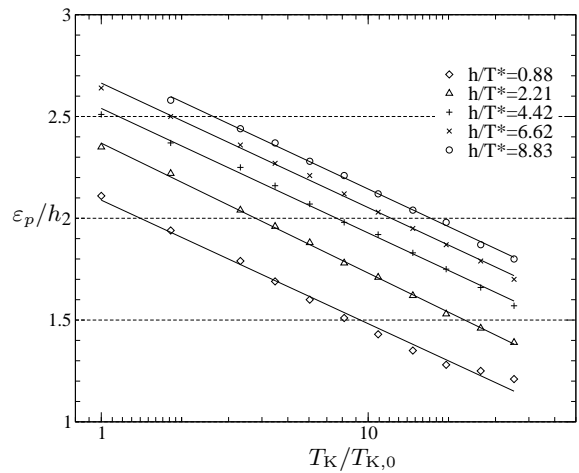}
\vskip-.2cm
\end{center} 
\caption{The dependence of the ratio ${\epsilon_p}(h)/h$  on $T_{\rm K}$ for various $h$. The energy
  scale $T^*$ is given by the corresponding values for $U/{\pi\Delta}=4$,
  which satisfies $T^*/T_{\mathrm{K},0}\approx 11$.} 
\label{logfit}
\end{figure}

We have also studied the dependence of the ratio $\epsilon_p(h)/h$ on the Kondo
temperature $T_{\rm K}$. In figure \ref{logfit} we plot the results starting from a
minimal $T_{\mathrm{K},0}$ corresponding to $ U/\pi\Delta=6$ for various
fields $h$. One finds that with decreasing Kondo temperature, which translates
to increasing renormalization of the quasiparticles, the peak 
splitting and thus the sensitivity towards the exposure to a magnetic field is
enhanced. The results can be shown to be fitted well by a logarithmic law, 
\begin{equation}
   \label{logfiteq}
  \frac{\epsilon_p(h,T_\mathrm{K})}h=
  \frac{\epsilon_p(h,T_{\mathrm{K},0})}h+a\ln\left(\frac{T_{\mathrm{K},0}}{T_{\rm 
  K}}\right).
\end{equation}
For all the curves one finds $a= 0.26\pm 0.01$ independent of $h$. 

\begin{figure}[!htb]
\begin{center}
\psfrag{w}{$\omega/T^*$}
\psfrag{rho}{$\pi\Delta\rho_{\mathrm{d}}(\omega)$}
\includegraphics[width=0.42\textwidth]{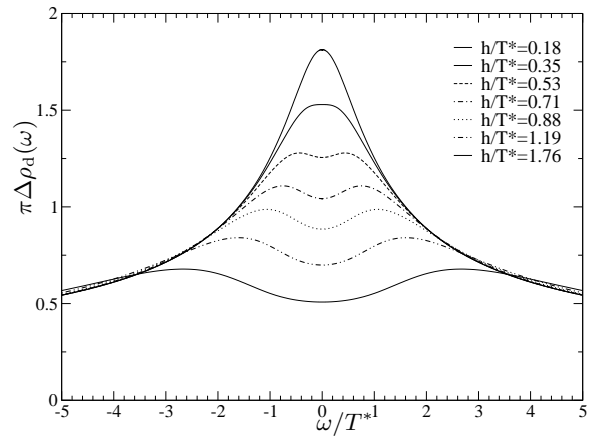}
\vskip-.3cm
\end{center} 
\caption{Total spectral density of the d-site $\rho_{\mathrm{d}}(\omega)$ for various
  fields $h$. One can see that the peak splitting becomes visible only for
  fields $h\gtrsim 0.5 T^*$.}
\label{totspectra}
\end{figure}

\noindent
For the total spectral density, which we denote by
$\rho_{\mathrm{d}}(\omega)$, we need to  
include the contribution from the down spin, which has its peak at $\epsilon_p(h)$,
$\rho_{\mathrm{d}}=\rho_{\mathrm{d},\uparrow}+\rho_{\mathrm{d}, 
\downarrow}$. The splitting between the up and down peaks in the total
spectrum is then $2\epsilon_p(h) f_c(h)$, where $f_c(h)$ is a correction factor due
to the overlap of the resonances.\cite{HBO05pre} In general it has to be
determined numerically, but for free quasiparticles (see equation (\ref{qpgfct})
without $\tilde\Sigma_{\sigma}$) it is given by
\begin{equation}
f_c(h)^2=1-\left(1-\left[1+\left(\tilde\Delta(h)/\tilde\epsilon_{\rm d}(h)
        \right)^2\right]^{1/2}\right)^2.
\label{fc}
\end{equation} 
For higher fields one has $f_c(\tilde\eta  
b,\tilde\Delta)\simeq 1-{\tilde\Delta^4}/{8(\tilde\eta  b)^4}$.
In figure \ref{totspectra} we give an example for the total spectral density for the
earlier used parameters $U/\pi\Delta=4$ and a range of magnetic
fields. Clearly, the peak splits above a critical field, $h\gtrsim 0.5 T^*$,
which is in agreement with results of Costi for the Kondo model\cite{Cos00}.

\subsection*{Comparison with experiments on quantum dots}

In section V.B we have calculated the lowest order temperature dependence of
the conductance through a quantum dot in a magnetic field, (\ref{Gtemp2}). 
This temperature dependence has been observed in the zero magnetic field
case\cite{GKSMM98}. We noted earlier that there is a sign change in this leading
temperature dependence at a values of the magnetic field $0<h<T^*$. The sign
change in the second term in equation (\ref{Gtemp}) occurs when $\rho_{\rm d}$
changes from a local maximum to a minimum as can be seen in figure \ref{totspectra}.
A qualitative explanation of this sign change is that the local spectral density on
at the Fermi level is suppressed with increasing magnetic field. At higher fields
when the spectral density develops two peaks then there are more thermally excited
states which can contribute to an increase of the conductance. 
This temperature dependence might be experimentally observable, since
estimates of the Kondo temperature are of the order $300mK$ 
corresponding to magnetic fields in the experimental range.\cite{KAGGKS04}
A difficulty might be that the overall response is reduced by the
$\cos^2(\pi m(h))$ factor in equation (\ref{Gtemp2}).

\begin{figure}[htbp]
  \centering
  \includegraphics[width=0.42\textwidth]{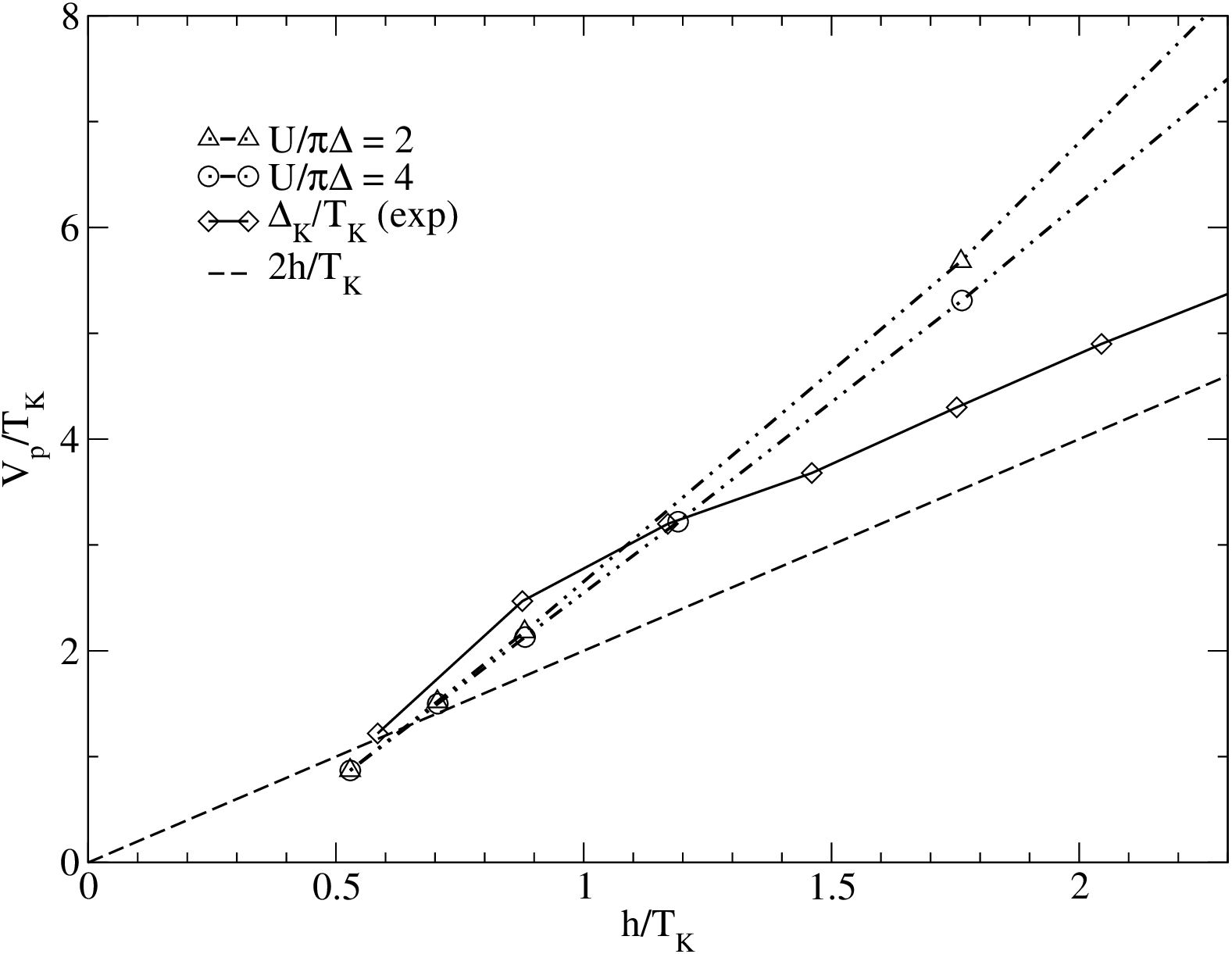}
  \caption{Peak position $V_p$ of the theoretical results from
    (\ref{didv}) obtained from the NRG spectra in comparison with experimental
    result $\Delta_{\rm K}$ (referred to as half the Kondo
    splitting)\cite{KAGGKS04}. The experimental Kondo temperature is inferred 
    from the critical field for peak splitting $B^{\rm (exp)}_c\approx 2{\rm T}$ and the
    result $h_c\approx 0.5842T_{\rm K}$  for strong coupling.\cite{HBO05pre} A dashed
    line corresponding to twice the Zeeman energy has been included for comparison.
    Note that, although all curves lie above $2h$, due to
    $\omega=V/2$ in the argument of $\rho_{\rm d}$ in equation (\ref{didv})
    the actual peak splitting does not exceed twice the Zeeman splitting in
    all cases.\cite{commentksplit}}
\label{splitcomp}
\end{figure}

Some of the quantum dot experiments\cite{KAGGKS04,AGKK05} in the presence of a magnetic field
have been performed in non-equilibrium situations with a finite source-drain
voltage $V$. Two peaks can be observed in the differential conductance
as a function of the voltage $V$ for fields strengths larger than a critical
value. There have been several interpretations\cite{MW00,LD01} of these results based on
the approximation of using the equilibrium Green function to evaluate
(\ref{current2}). With this approximation at $T=0$ we get an expression for
the differential conductance $G(V)$ as a function of the voltage $V$,
\begin{equation}
 G(V)=\frac{dI}{dV}=\frac{G_0\pi\Delta}{2} \rho_{\mathrm{d}}(eV/2).
\label{didv}
\end{equation} 
In this approximation 
$G(V)$ is  directly proportional to the total equilibrium spectral density 
evaluated at $\omega=eV/2$, which is shown in figure \ref{totspectra}.
To test whether the experimental results can be explained on the basis of
equation (\ref{didv}) we have extracted the magnitude of the peak splitting
for $U/\pi\Delta=2,4$  and compared it with experimental
results\cite{KAGGKS04}, which are displayed in figure \ref{splitcomp}.  

We can see in figure \ref{splitcomp} that, whilst there is an agreement in the
range  $h/T_{\rm K}\simeq 1$, in general there does not appear to be a satisfactory
explanation of the experimental results based on approximating  the non-equilibrium Green
function by the equilibrium one. 
Hewson, Bauer and Oguri\cite{HBO05pre} have calculated the non-equilibrium
Green
function for a small but finite voltage $V$, and find that a finite voltage reduces the peak position 
substantially. We conclude that an accurate agreement of experiment and theory rests therefore on an
accurate description of the steady state situation out of equilibrium.
We are currently working on an extension of renormalized perturbation theory in the
non-equilibrium formalism, which takes this situation into account appropriately.

\section{Summary}
We have shown that in an arbitrary magnetic field $H$ it is possible
to describe the  quasiparticles of a Fermi liquid regime by field dependent
parameters. We have used 
the particle-hole symmetric Anderson model as an illustration because of its
wide range of applications and because it serves as a local model of strong
correlation physics. For this model the three relevant field dependent
parameters are the renormalized impurity levels with spin $\sigma$,
$\tilde\epsilon_{\mathrm{d},\sigma}(H)$, the  quasiparticle resonance width,
$\tilde\Delta(H)$ and the local interaction $\tilde U(H)$. We have shown
how these can be deduced from NRG calculations of the low lying
excitations. Once the renormalized parameters are known, the impurity
spin and charge susceptibility, the specific heat coefficient and the
induced impurity  magnetization at $T=0$ for arbitrary magnetic field can be
expressed exactly by substituting into the relevant formulae derived from a renormalized
perturbation theory. The leading temperature dependent corrections to the
susceptibility and magnetization can be obtained in a similar way. We have
also extended the renormalized perturbation expansion in order to calculate
the leading temperature dependence for the finite conductivity 
due to scattering from an impurity in a metallic host, and for the conductance
through a quantum dot.

By choosing the bare parameters in the absence of the field to correspond to
the strongly correlated or Kondo limit, the de-renormalization of the
quasiparticles can be followed as the magnetic field strength  is increased from
zero. For extremely large magnetic fields the parameters revert to their
bare values.
This approach gives an overview of the low energy behaviour of the model as a
function of the applied magnetic field strength.\\
A number of physical properties are found to change qualitatively  in the
strongly correlated case for magnetic field strengths in the range
$0<g\mu_{\rm B}H<T_{\rm K}$, where $T_{\rm K}$ is the Kondo  
temperature. This should be a physically accessible magnetic
field range for many systems. The $T^2$ coefficient of the magnetic
susceptibility, the conductivity from a magnetic impurity in the strong
correlation regime, and the conductance through a quantum dot all change sign
in this magnetic field range.

The approach developed here is a general one and is equally applicable to
asymmetric impurity models and to lattice models. For lattice models, for which
dynamical mean field theory is applicable, 
similar NRG methods to those employed here can be used\cite{KHE05pre}, and our
calculations are currently being 
extended to models of heavy fermions. We note that the approach is not
restricted to the NRG method, the relevant renormalized parameters
could also be estimated using other theoretical techniques, variational
methods for example.

\bigskip
\noindent{\bf Acknowledgement}\par
\noindent
We wish to thank the EPSRC (Grant GR/S18571/01) for
financial support,  A. Oguri  and  D. Meyer  for helpful
discussions on many aspects of the work described here. One of us (J.B.)
thanks the Gottlieb Daimler-and Karl Benz Foundation for financial support.

\par
\section{Appendix: The Renormalized Perturbation Approach} 
We give a brief synopsis of the renormalized perturbation approach.
This approach is best developed in the field theoretical
Lagrangian formalism where the renormalization of the field is a more natural
concept. The  Lagrangian corresponding to the bare Anderson  ${\cal L}_{\rm
AM}(\epsilon_{\mathrm{d}},\Delta,U)$ can be rewritten in  the form
\begin{equation}
{\cal L}_{\rm AM}(\epsilon_{\mathrm{d},\sigma},
\Delta,U)={ \cal L}_{\rm AM}(\tilde\epsilon_{\mathrm{d},\sigma},
\tilde\Delta,\tilde U)+ {\cal L}_{\rm
  ct}(\lambda_1,\lambda_2,\lambda_3),\label{rlag}
\end{equation}
where the renormalized parameters,   $\tilde\epsilon_{\mathrm{d},\sigma}$ and 
$\tilde\Delta_{\sigma}$, are defined in terms of the self-energy
$\Sigma_{\sigma}(\omega)$ of the one-electron Green function for the impurity state,
\begin{equation}
G_{\mathrm{d},\sigma}(\omega)={1\over
    \omega-\epsilon_{\mathrm{d}\sigma}+i\Delta-\Sigma_\sigma(\omega)},
\label{gf}
\end{equation}
and are given by 
\begin{equation}
\tilde\epsilon_{\mathrm{d},{\sigma}}=z_{\sigma}(\epsilon_{\mathrm{d},{\sigma}} 
+\Sigma_\sigma(0)),\quad\tilde\Delta_{\sigma} =z_\sigma\Delta,
\label{ren1}
\end{equation} 
where $z_{\sigma}$ is given by
$z_{\sigma}={1/{(1-\Sigma_{\sigma}'(0))}}$.
The renormalized or quasiparticle interaction  $\tilde U$, is defined in terms
of the local irreducible 4-vertex
$\Gamma_{\uparrow\downarrow}(\omega_1,\omega_2,\omega_3,\omega_4)$ at zero frequency,
  \begin{equation} 
\tilde U=z_{\uparrow}z_{\downarrow}\Gamma_{\uparrow\downarrow}(0,0,0,0).
\label{ren2}\end{equation}
The remaining part of the Lagrangian ${\cal L}_{\rm
  ct}(\lambda_1,\lambda_2,\lambda_3)$ 
contains three counter terms with coefficients $\lambda_1$, $\lambda_2$ and $\lambda_3$. 
A perturbation expansion in powers of the quasiparticle interaction $\tilde U$
  can then be carried out provided the counter terms are also taken into
  account.  The counter terms $\lambda_1$, $\lambda_2$ and $\lambda_3$ are
  determined to each order in $\tilde U$ by the renormalization conditions\cite{Hew93},
\begin{equation}\tilde\Sigma_\sigma(0)=0,\quad \tilde\Sigma'_\sigma(0)=0,\quad
  \tilde\Gamma_{\uparrow,\downarrow}(0,0,0,0)=\tilde U \label{rcond}\end{equation}
where $\tilde\Sigma(\omega)$ and
$\tilde\Gamma_{\uparrow\downarrow}(\omega_1,\omega_2,\omega_3,\omega_4)$
are the renormalized self-energy and irreducible 4-vertex, respectively.
\par
The renormalized parameters, as defined in equations (\ref{ren1}) and
(\ref{ren2}), can be calculated in low order perturbation theory \cite{Hew01},
but these calculations cannot be extended to the more physically interesting
strong coupling regime $U/\pi\Delta>2$. 
These  parameters, however, can be identified with those in equation (\ref{rham}), 
and so they can alternatively be deduced from the levels in  the numerical
renormalization group calculations, which can be carried out for all values of $U$.
The expressions given earlier for the impurity occupation number at $T=0$, the
specific heat coefficient and spin and charge susceptibilities, (\ref{qpfsr}),
(\ref{rgam}) and (\ref{rsus}), 
are exact and  correspond to
first order calculations in a renormalized perturbation expansion\cite{Hew93} in powers of
$\tilde U$.  Calculations carried out to second order in $\tilde U$
give exact results for the leading low order temperature dependence to the
impurity resistivity \cite{Hew93}  and the leading non-linear term in the differential 
conductivity \cite{Ogu01,Ogu05}.\par

\bibliography{artikel,biblio1,footnote}

\begin{thebibliography}{38}
\bibitem[{com({\natexlab{a}})}]{comment1}
\bibinfo{note}{This is a corrected result from an earlier treatment given in
  reference\cite{Hew05}, where cross-terms leading to contributions from the
  real part of the self-energy were not included}.
\bibitem[{com({\natexlab{b}})}]{comment2}
\bibinfo{note}{Due to the symmetry of the setup it is reasonable to assume that
  when a bias voltage $V$ is applied the chemical potential on the d-site is
  shifted to the middle of $\mu_L$ and $\mu_R$}.
\bibitem[{com({\natexlab{c}})}]{commentksplit}
\bibinfo{note}{When a finite voltage is applied there is an average shift of
  $V/2$ of the levels on the dot.\cite{HBO05pre} This shift does not appear to
  have been taken into account in the interpretation of the experiments. If
  this shift is included it leads to an estimate of the splitting of the Kondo
  resonance, which is half the value reported\cite{KAGGKS04}}.
\end{thebibliography}

\end{document}